\documentclass[fleqn,usenatbib]{mnras}

\usepackage{newtxtext,newtxmath}
\usepackage{multirow}

\usepackage[T1]{fontenc}

\DeclareRobustCommand{\VAN}[3]{#2}
\let\VANthebibliography\thebibliography
\def\thebibliography{\DeclareRobustCommand{\VAN}[3]{##3}\VANthebibliography}

\usepackage{graphicx}	
\usepackage{amsmath}	

\graphicspath{ {./Thesis/Thesis_Figures/} {./MG_MSW/} {./SG_MSW/} }

\title[RSG Shock Cooling Model: II]{Shock cooling emission from explosions of red super-giants: II. An analytic model of deviations from blackbody emission}

\author[Morag et al.]{
Jonathan Morag\thanks{E-mail: jonathan.morag@weizmann.ac.il},$^{1}$
Ido Irani,$^{1}$
Nir Sapir,$^{1,2}$
Eli Waxman$^{1}$
\\
$^{1}$Weizmann Institute of Science, Rehovot, Israel\\
$^{2}$Soreq Nuclear Center, Yavne, Israel\\
}

\date{Accepted XXX. Received YYY; in original form ZZZ}

\pubyear{2022}

\begin{document}
\label{firstpage}
\pagerange{\pageref{firstpage}--\pageref{lastpage}}
\maketitle

\begin{abstract}

Light emission in the first hours and days following core-collapse supernovae (SNe) is dominated by the escape of photons from the expanding shock heated envelope. In a preceding paper, Paper I, we provided a simple analytic description of the time dependent luminosity, $L$, and color temperature, $T_{\rm col}$, valid up to H recombination ($T\approx0.7$~eV), for explosions of red supergiants with convective polytropic envelopes without significant circum-stellar medium (CSM). The analytic description was calibrated against "gray" (frequency-independent) photon diffusion numeric calculations. Here we present the results of a large set of 1D multi-group (frequency-dependent) calculations, for a wide range of progenitor parameters (mass, radius, core/envelope mass ratios, metalicity) and explosion energies, using opacity tables that we constructed (and made publicly available), including the contributions of bound-bound and bound-free transitions. We provide an analytic description of the small, $\simeq10\%$ deviations of the spectrum from blackbody at low frequencies, $h\nu< 3T_{\rm col}$, and an improved (over Paper I) description of `line dampening' for $h\nu> 3T_{\rm col}$. We show that the effects of deviations from initial polytropic density distribution are small, and so are the effects of `expansion opacity' and deviations from LTE ionization and excitation (within our model assumptions). 
A recent study of a large set of type II SN observations finds that our model accounts well for the early multi-band data of more than 50\% of observed SNe (the others are likely affected by thick CSM), enabling the inference of progenitor properties, explosion velocity, and relative extinction.
\end{abstract}

\begin{keywords}
radiation: dynamics – shock waves – supernovae: general
\end{keywords}



\section{Introduction}
In core collapse supernovae (SNe) explosions, a radiation mediated shock (RMS) traverses outwards through the stellar progenitor, heating and expelling material as it passes. If no significant circumstellar material (CSM) is present around the star, arrival of the shock at the surface produces a hard UV/X-ray $\sim10^{45} \, \rm erg \, s^{-1}$ `shock-breakout' emission, lasting from tens of minutes to an hour. The breakout pulse is then followed in the coming hours and days by thermal UV/optical `shock-cooling' emission, caused by diffusion of photons out of the shock-heated stellar ejecta. Typical luminosities and temperatures during shock-cooling are of the order $10^{42}-10^{44} \rm \, erg \, s^{-1}$, and $1-10$ eV. As the photons diffuse out, deeper parts of the ejecta are gradually exposed over time \citep[see][for reviews]{waxman_shock_2016,levinson_physics_2020}.

In order to constrain the properties of the progenitor star, it is helpful to have high cadence multi-band observations in the first hours of shock-cooling \citep[see][for a detailed discussion of the theoretical considerations and observational status]{morag_shock_2023}. Among these measurements, ultraviolet observations are especially important, as they are closer to the thermal emission peak, and can be used to determine the emission temperature and the UV extinction self-consistently \citep{rabinak_early_2011,sapir_uv/optical_2017,sagiv_science_2014,rubin_exploring_2017}.
Combined with an accurate theoretical model, these measurements can be used to reproduce the progenitor and explosion parameters, including radius, surface composition, explosion energy per unit mass, and the extinction. Analytic models are especially important for solving the "inverse problem" of inferring system parameters and uncertainties from the observed spectral energy distribution (SED).

Emission during shock-cooling is amenable to modeling in the case of `envelope breakout' (i.e. in the absence of CSM with signifcant optical depth), largely because the system is near local thermal equilibrium (LTE) at this time. However, catching supernovae within the first hour presents a practical challenge, and few such observations have been achieved  \citep[see][for a representative list]{morag_shock_2023}. Existing and upcoming observatories, such as the Zwicky Transient Factory (ZTF) \citep{gal-yam_real-time_2011}, the upcoming Vera Rubin Observatory \citep{ivezic_lsst_2019}, and the expected launch of the wide-field UV space telescope ULTRASAT \citep{sagiv_science_2014,shvartzvald_ultrasat_2023} will greatly increase the quantity and quality of early measurements, enabling a systematic study.

The numeric calculations presented here provide a systematic analysis of the deviation of the spectra of the emitted radiation from blackbody, which enables an improvement of the accuracy of the analytic models. We adopt a `multigroup' (MG) numeric approach, where radiative transfer is solved under the diffusion approximation
including coupling to hydrodynamics. Several frequency-dependent codes using different approximations and schemes have been used to address the shock breakout and cooling problem: STELLA \citep{blinnikov_stella_2011} is a 1 dimensional code that solves for the angle-dependent intensity with a variable Eddington factor, employing a ray-tracing scheme. Their opacity includes free-free, bound-free and atomic lines from \citet{kurucz_atomic_1995}, and employs the Sobolev approximation, based on \citet{Eastman_Spectrum_1993}. STELLA should contain the required physical components for modeling the SED of shock-cooling emission, and this has been done in several works \citep{blinnikov_comparative_1998,blinnikov_radiation_2000,tominaga_shock_2011,tominaga_properties_2009,forster_delay_2018}. \citet{sapir_numeric_2014} numerically solved the planar shock breakout problem under the diffusion approximation and incorporating free-free opacity. They also include inelastic Compton scattering, which is important for shock breakouts at high velocities. Other hydrodynamically coupled multigroup codes are presented in the literature \citep{vaytet_numerical_2011,skartlien_multigroup_2000,tominaga_time-dependent_2015}, though they have not been used or formulated to solve for shock-cooling emission.

Many other codes focus computational efforts on line modeling and are specialized to describe emission at later phases of the supernova, when the ejecta is freely-coasting. The ARTIS \citep{kromer_time-dependent_2009} and SEDONA \citep{roth_monte_2015} codes both solve frequency-dependent, time-dependent radiative transfer using Monte Carlo. ARTIS is primarily used to solve type Ia SN problems, while SEDONA has also been used to calculate core-collapse supernova emission \citep[e.g.][]{kleiser_fast_2012,tsang_comparing_2020,jacobson-galan_circumstellar_2022}. The latter contains a module for coupling hydrodynamics to radiation in a Lagrangian sense, though to our knowledge this has not been used in the context of the first day of shock-cooling emission. Other multigroup codes solve for radiative transfer using static "snapshots" of the hydrodynamic profiles, under the assumption that the photon diffusion time is short compared to the dynamical time of the ejecta
\citep[][]{baron_preliminary_2000,dessart_numerical_2015,baron_determination_2003,kerzendorf_spectral_2014,ergon_monte-carlo_2018}. 
It is not guaranteed that this approximation is valid at the earliest times, though our results suggest that it may be reasonable close to recombination if one employs realistic density and temperature profiles.

\defcitealias{sapir_non-relativistic_2011}{SKW I}
\defcitealias{katz_non-relativistic_2012}{KSW II}
\defcitealias{sapir_non-relativistic_2013}{SKW III}
\defcitealias{rabinak_early_2011}{RW11}
\defcitealias{sapir_uv/optical_2017}{SW17}
\defcitealias{nakar_early_2010}{NS10}
\defcitealias{shussman_type_2016}{SWN16}
\defcitealias{piro_shock_2021}{PHY21}
\defcitealias{morag_shock_2023}{Paper I}

In a preceding paper, \citet[][hereafter \citetalias{morag_shock_2023}]{morag_shock_2023}, we modeled shock-cooling emission following breakout from a spherical envelope, assuming local thermal equilibrium (LTE). Using earlier analytic results \citep[][hereafter \citetalias{sapir_non-relativistic_2011}, \citetalias{katz_non-relativistic_2012}, \citetalias{sapir_non-relativistic_2013}, \citetalias{rabinak_early_2011}, \citetalias{sapir_uv/optical_2017}]{sapir_non-relativistic_2011,katz_non-relativistic_2012,sapir_non-relativistic_2013,rabinak_early_2011,sapir_uv/optical_2017}, we derived an analytic description of the bolometric luminosity $L$ and color temperature $T_{\rm col}$, that provides a good approximation (to 10\% in $L$ and 5\% in $T_{\rm col}$) of the results of our `gray' diffusion simulations for the emission over $\sim1$ hour to $\sim1$ week after shock breakout for a wide range of explosion energies and progenitor parameters (radii, core and envelope masses, metallicities). The analytic expression is advantageous relative to those of previous analytic works \citep[\citetalias{sapir_non-relativistic_2011,katz_non-relativistic_2012,sapir_non-relativistic_2013,rabinak_early_2011};][]{nakar_early_2010,shussman_type_2016,piro_numerically_2017} thanks to its calibration using a large set of numeric results  incorporating a realistic opacity with free-free, bound-free, and bound-bound components, which have an important effect on emission. We also used a set of MG diffusion simulations to show that the spectrum is well described by a Planck spectrum with an effective color temperature, except at UV frequencies where the flux may be significantly suppressed due to line absorption.

In this work we relax the assumption that the photons are at LTE and account for frequency-dependent emission, absorption and transport both numerically and analytically. We produce hydrodynamically coupled 1-dimensional multigroup simulations for a wide range of explosion energies and progenitor parameters that span the parameter range of Paper I - explosion energies in the range $E=10^{50}-10^{52}$~erg, progenitor masses and radii $M=2-40M_{\odot}$,  $R=3\times10^{12}-2\times10^{14}$~cm, envelope to core mass ratios $M_{\rm env}/M_{\rm c}=10-0.3$, and metallicities $Z=0.1Z_{\odot}-Z_{\odot}$. Our simulations are based on the work of \citet{sapir_numeric_2014}. Their code allows inelastic Compton scattering, which we do not incorporate in the bulk of the simulations, due to its negligible effect during shock-cooling. We include a frequency-dependent opacity $\kappa_{\nu}$, for which we use both the publicly available TOPS code \citep{colgan_new_2016} and one that we developed ourselves and is now available to the community \cite{morag_frequency_2023}. Relative to TOPS, our code is largely open source, and can produce opacity tables at arbitrary density, temperature, and frequency resolution. It is based on experimentally verified atomic line lists \citep{kurucz_atomic_1995}. We note that our MG simulations cannot be used to make predictions for the exact spectrum including lines, as they have finite frequency resolution, and do not include a treatment of "expansion opacity" effects and of deviations of excitation and ionization states from LTE (see \S~\ref{sec: finite frequency resolution}). However, they are useful for predicting the coarser-grained, line averaged SED in shock-cooling following envelope breakout. 

A recent study of a large set of type II SN observations \citep{irani_sn_2022} finds that our model accounts well for the early multi-band data of 50\% of observed SNe, corresponding to 70\% of the intrinsic SN distribution (the others are likely affected by thick CSM). This agreement enables the inference of progenitor properties, explosion velocity, and relative extinction by using the formulae provided in this paper.

This paper is organized as follows. In \S~\ref{sec: Physics of RMS and SCE} we define our notation and summarize the analytic results of \citetalias{sapir_non-relativistic_2011}, \citetalias{katz_non-relativistic_2012}, \citetalias{rabinak_early_2011}, \citetalias{sapir_uv/optical_2017} and \citetalias{morag_shock_2023}, that we use in this paper. In \S~\ref{sec:Code description} we describe our numeric calculations and present convergence and code verification tests. In \S~\ref{sec:kappa} we describe our opacity tables, as well as convergence and verification tests for these. In \S~\ref{sec:analytic formula} we derive an analytic description of the deviations of shock cooling emission spectra from blackbody, and in \S~\ref{sec:numeric_res} we compare the analytic description to simulation results. In sections \S~\ref{sec: finite frequency resolution} and \S~\ref{sec:NLTE}, we address the sensitivity of our results to "expansion opacity" corrections and to deviations of plasma ionization and excitation from those of LTE. A comparison to earlier work, including to STELLA radiation transport calculations for several non-polytropic profiles obtained using MESA stellar evolution calculations \citep{blinnikov_comparative_1998,tominaga_shock_2011,kozyreva_shock_2020}, is given in \S~\ref{sec: Previous works}. The agreement of our results with these earlier calculations provides additional support to our code's validity, and to the conclusion of the detailed analysis of SW17, who demonstrated that the shock cooling emission is not sensitive to deviations of the density profile from a polytropic one. Our results are summarized and discussed in \S~\ref{sec: Summary}.

\section{RMS breakout and Shock-Cooling Emission- summary of earlier analytic results used in this paper}
\label{sec: Physics of RMS and SCE}

RMS breakout and shock-cooling is extensively discussed in the literature and briefly summarized in \citetalias{morag_shock_2023}. At radii $r$ close to the stellar radius $R$ ($\delta \equiv (R - r)/R\ll 1$), the initial density of a polytropic envelope approaches a power-law, 
\begin{equation}
\label{eq:rho_in}
    \rho_0 = f_\rho \bar{\rho} \delta^n.
\end{equation}
Here $\bar{\rho}\equiv M/(4\pi R^3/3)$ is the average pre-explosion density of the ejecta (exculding the mass of a possible remnant), and $n=3/2$ for convective RSG envelopes. $f_\rho$ is a numerical factor, of order unity for convective envelopes, that depends on the inner envelope structure \citep{matzner_expulsion_1999,calzavara_supernova_2004,sapir_uv/optical_2017}. The predicted breakout and cooling emission are nearly independent of $f_\rho$.

As the shock approaches the edge of the star, it accelerates down the steep density profile and the flow approaches the self-similar solutions of \citet{gandelman_shock_1956,sakurai_problem_1960}. The shock velocity diverges in this regime as
\begin{equation}
\label{eq:vs}
    \rm v_{\rm sh} = v_{\rm s\ast} \delta^{-\beta_1 n},
\end{equation}
with $\beta_1=0.19$, and with $\rm v_{\rm s\ast}$ a constant defined by eq.~(\ref{eq:vs}). Based on numerical calculations, \citet{matzner_expulsion_1999} find 
\begin{equation}
\label{eq:vstar}
    {\rm v_{\rm s\ast}}\approx 1.05 f_\rho^{-\beta_1}{\rm v_\ast,\quad v_\ast}\equiv\sqrt{E/M},
\end{equation}
where $M$ is the mass of the ejecta, $E$ is the energy deposited in the ejecta, and $\rm v_\ast$ is its characteristic expansion velocity. This approximation holds to better than 10\% for $M_{\rm env}/M_{\rm c}<1/3$, and overestimates $\rm v_{\rm s\ast}$ by approx.~20\% for $M_{\rm env}/M_{\rm c}=0.1$ \citepalias[see figure 7 of ][]{sapir_uv/optical_2017}.

Breakout takes place when the scattering optical depth of the plasma layer lying ahead of the shock equals $\tau_{\rm es}=c/\rm v_{\rm sh}$ \citep{ohyama_explosion_1963} \footnote{Scattering strongly dominates absorption in the relevant highly ionized regime.}. We denote the shock velocity $\rm v_{\rm sh}$ and pre-shock envelope density $\rho_0$ at this point in terms of breakout parameters $\rho_{\rm bo}$ and $\rm v_{\rm bo}$, respectively. We may rewrite eqs.~(\ref{eq:rho_in}) and~(\ref{eq:vs}) as
\begin{equation}\label{eq:rho_v_0_def}
  \rho_0=\rho_{\rm bo} ({\rm v_{\rm bo}}\tau/c)^{n/(1+n)},\quad
  {\rm v_{\rm sh}} = {\rm v_{\rm bo} (v_{\rm bo}}\tau/c)^{-\beta_1 n/(1+n)}.
\end{equation}
The location at which breakout "occurs", i.e. where $\tau=c/\rm v_{\rm sh}$, is given by
\begin{equation} \label{eq:delta_bo_def}
    \delta_{\rm bo} = (n+1)\frac{c }{ \kappa \rho_{\rm bo} {\rm v_{\rm bo}} R},
\end{equation}
where $\kappa$ is the opacity.

For RSGs, $\rho_{\rm bo}$ and ${\rm v_{\rm bo}}$ are approximately related to the progenitor parameters and explosion energy by \citep{waxman_shock_2016}
\begin{align} \label{eq:rho_v_0_approx}
  \rho_{\rm bo} & = 1.16 \times 10^{-9} M_{0}^{0.32} {\rm v_{\ast,8.5}}^{-0.68} R_{13}^{-1.64} \kappa_{0.34}^{-0.68} f_{\rho}^{0.45}\, \rm g \, cm^{-3}, \nonumber\\
  {\rm v_{\rm bo}/v_{\ast}} & = 3.31 M_{0}^{0.13} {\rm v_{\ast, 8.5}}^{0.13} R_{13}^{-0.26} \kappa_{0.34}^{0.13} f_{\rho}^{-0.09}.
\end{align}
Here, $R= 10^{13}R_{13}$~cm, $\kappa=0.34 \kappa_{0.34} \rm \, cm^2 g^{-1}$, ${\rm v_\ast=v_{\ast,8.5}} 10^{8.5} \, \rm cm s^{-1}$, and $M=1 M_{0} M_\odot$. The duration over which the breakout pulse $t_{\rm bo}$ is emitted from the star is approximately given by the shock crossing time of the breakout layer $t_{\rm bo}$,
\begin{equation}
\label{eq:tbo}
   \frac{\delta_{\rm bo}R}{\rm v_{\rm bo}} =\frac{(n+1)c }{ \kappa \rho_{\rm bo} v_{\rm bo}^2}= (n+1)t_{\rm bo}=74.9\, \rho_{\rm bo,-9}^{-1}\kappa_{0.34}^{-1}{\rm v_{\rm bo,9}}^{-2}{\, \rm s},
\end{equation}
where $\rho_{\rm bo}=10^{-9}\rho_{\rm bo,-9}{\rm \, g \, cm^{-3}}$ and $\rm v_{\rm bo}= 10^{9} v_{\rm bo,9} \,cm\,s^{-1}$. The observed pulse duration may be longer than this intrinsic duration due to light travel time effects, which spread the pulse over $R/c$.
$\delta_{\rm bo}$ is given as a function of progenitor parameters and explosion energy as
\begin{equation}
\label{eq:dbo}
    \delta_{\rm bo}=0.02 \, R_{13}^{0.90} (f_\rho M_0 \, {\rm v_{\rm s*,8.5}}\, \kappa_{0.34})^{-0.45},
\end{equation}
where ${\rm v_{\rm s*}=v_{\rm s*,8.5}} 10^{8.5}$.

For later use, we provide here the density profile during the spherical phase, given by eq. (9) of \citetalias{rabinak_early_2011}. We recast this equation in terms of $r$ and $t$ using \citetalias{rabinak_early_2011} eqs.~(3), (4) and (8), and eq. (\ref{eq:vstar}) from this paper,
\begin{equation}
    \rho(r,t) = 1.69 \times 10^{-11}\, f_{\rm \rho} M_0 \, v_{s*,8.5}^{8.73} r_{14}^{-11.73} t_{\rm d}^{8.73} \, \rm g \, cm^{-3},
\label{eq:rho_rt_vsstar}
\end{equation}
where the radial coordinate is $r=r_{14} 10^{14}$ cm. Alternatively, using eq. \ref{eq:rho_v_0_def},
\begin{equation}
    \rho(r,t) = 1.82 \times 10^{-11} \, R_{13}^2 {\rm v_{\rm bo,9}^{7.73}} \kappa_{0.34}^{-1} r_{14}^{-11.73} t_{\rm d}^{8.73} \, \rm g \, cm^{-3}.
    \label{eq:rho_rt_bo}
\end{equation}
Similarly, assuming a self-similar diffusion profile \citep{chevalier_early_1992}, we have for the temperature:
\begin{equation}
    T(r,t) = 4.83 \, R_{13}^{1/4} (f_{\rho}M_0)^{0.27} {\rm v_{\rm s*,8.5}^{2.66}} \kappa_{0.34}^{0.02} t_{\rm d}^{2.14} r_{14}^{-3.18} \, \rm eV,
    \label{eq:T_rt_vsstar}
\end{equation}
\begin{equation}
    T(r,t) = 5.02 \, R_{13}^{0.71}  {\rm v_{\rm bo,9}^{2.31}} \rho_{\rm bo,9}^{-0.08} \kappa_{0.34}^{-1/3} t_{\rm d}^{2.14} r_{14}^{-3.18} \, \rm eV.
    \label{eq:T_rt_bo}
\end{equation}

In \citetalias{morag_shock_2023}, we described shock cooling emission by interpolating between the exact planar phase solution (\citetalias{sapir_non-relativistic_2011,katz_non-relativistic_2012,sapir_non-relativistic_2013}) valid at early times (hours), and the later approximate spherical phase solution (\citetalias{rabinak_early_2011,sapir_uv/optical_2017}).
The combined bolometric luminosity $L$ and emission (color) temperature $T_{\rm \, col}$ are given by
\begin{equation}
    L/L_{\rm br}=\tilde{t}^{-4/3}+\tilde{t}^{-0.172}\times A\exp\left(-\left[at/t_{\rm tr}\right]^{\alpha}\right),
    \label{eq:L_trans}
\end{equation}
\begin{equation}
    T_{\rm col}/T_{\rm col,br}=\min\left[0.97\,\tilde{t}^{-1/3},\tilde{t}^{-0.45}\right].
    \label{eq:T_trans}
\end{equation}
Assuming a blackbody spectral distribution, the emitted luminosity is then given by,
\begin{equation}
    L_{\rm BB}=L\times\pi B_{\nu}(T_{\rm col})/\sigma T_{\rm col}^{4}.
    \label{eq:L_nu_BB_formula}
\end{equation}
Here $\{A,a,\alpha\}=\{0.9,2,0.5\}$  \citepalias[note slight difference from ][]{sapir_uv/optical_2017}, and $\tilde{t}\equiv t / t_{\rm br}$. $t_{\rm tr}$ is roughly the time at which the photons will be able to diffuse out of the envelope in dynamical time. The br (break) subscript marks the values at the transition between the planar and spherical phase. They are given in terms of the model parameters $\rm v_{\rm s*}$\,,$f_{\rho}M$, and $R$ as
\begin{equation}
\label{eq:t_br_of_vs}
    t_{\rm br}= 0.86 \, R_{13}^{1.26} \rm v_{\rm s*,8.5}^{-1.13}
(f_{\rho}M_0\kappa_{0.34})^{-0.13}\,\text{hrs},
\end{equation}
\begin{equation}
\label{eq:L_br_of_vs}
L_{\rm br}=3.69\times10^{42} \, R_{13}^{0.78} \rm v_{\rm s*,8.5}^{2.11}
(f_{\rho}M_0)^{0.11} \kappa_{0.34}^{-0.89}\,{\rm erg \, s^{-1}},
\end{equation}
\begin{equation}
\label{eq:T_br_of_vs}
T_{\rm col,br}= 8.19 \, R_{13}^{-0.32} \rm v_{\rm s*,8.5}^{0.58}
(f_{\rho}M_0)^{0.03} \kappa_{0.34}^{-0.22}\,{\rm eV}.
\end{equation}
$M_0$ denotes mass in units of solar mass. Both the break values and $t_{\rm tr}$ can be directly deduced from observations.

Eqs. (\ref{eq:L_trans})-(\ref{eq:L_nu_BB_formula}) are valid at times $\max \left[ t_{\rm bo}, 3R/c \right]< t < \min \left[ t_{0.7 \rm \, eV} , t_{\rm tr}/a \right]$ (note that \citetalias{morag_shock_2023} did not explicitly include the shock-breakout early validity time $t_{\rm bo}$), where
\begin{equation}
    3 R / c = 0.67 \, t_{\rm br,3}^{-0.1} \,L_{br,42.5}^{0.55} \,T_{br,5}^{-2.21} \,\text{hrs},
    \label{eq:3Rc 1st time}
\end{equation}
\begin{equation}
    t_{0.7 \, \rm eV} = 8.01 \,
    t_{\rm br,3} \, T_{\rm br,5}^{2.22} \,
    \text{days}.
    \label{eq:t07eV 1st time}
\end{equation}
The former is the time past which we showed light-travel time effects to be unimportant. The latter is the time at which the photosphere temperature is $T=0.7$ eV, based on \citetalias{rabinak_early_2011}, roughly corresponding to Hydrogen recombination. The transparency time, $t_{\rm tr}$\footnote{Our $t_{0.7 \, \rm eV}$ and $t_{\rm tr}$ correspond roughly to $t_2$ and $t_4$ (or $t_5$) of \citet{utrobin_optimal_2007}.}, occurs roughly when the dynamical time matches the diffusion time, given by \citetalias{sapir_uv/optical_2017}
\begin{equation}
\label{eq:t_transp}
    \begin{split}
        \begin{aligned}
            t_{\rm tr} &= \sqrt{\frac{\kappa M_{\rm env}} {8 \pi c v_{\rm s \ast} }},  \\ &= 19.5 \, \sqrt{M_{\rm env,0} \kappa_{0.34} v_{\rm s*,8.5}^{-1}} \, \text{days}.
        \end{aligned}
    \end{split}
\end{equation}
For later use we also define the homologous time $t_{\rm hom}$, which is also the early validity of the \citetalias{rabinak_early_2011,sapir_uv/optical_2017} formula:
\begin{equation}
    t_{\rm hom} = R/5v_{\rm s,*} = 0.1 R_{13}/v_{\rm s*,8.5} \rm \,  days.
    \label{t_homologous}
\end{equation}

During shock-cooling, the luminosity is determined at the diffusion depth, the location from which photons will diffuse outwards in dynamical time \citep[see e.g. ][]{rabinak_early_2011}. The color temperature meanwhile, is roughly determined by the temperature at the thermal depth, the last absorption surface for diffusing photons (the radius from which photons diffuse out of the ejecta without further absorption). In this paper we treat these quantities as frequency-dependent (compare with the approximate prescription, \citetalias{morag_shock_2023}, eq. 30). Specifically the thermal depth $r_{\rm col}$ is defined by
\begin{equation}
\tau_{\star,\nu}(r=r_{{\rm col},\nu})\equiv\int_{r_{{\rm col},\nu}}^{\infty}\rho\sqrt{3\kappa_{\rm abs,\nu}\left(\kappa_{\rm abs,\nu}+\kappa_{\rm es}\right)}dr'=1,
    \label{eq: Thermal Depth Integral k_abs_nu}
\end{equation}
where the abs, es and $\nu$ subscripts indicate absorption, (electron) scattering and frequency dependence, respectively. This integral is often approximated in the literature as $\sqrt{3\tau_{\rm abs}\tau_{\rm es}}$
or $\sqrt{3\tau_{\rm abs}(\tau_{\rm abs}+\tau_{\rm es})}$. We find that the choice of approximation has a negligible effect on the SED due to the steep $\rho(r)$ dependence,
with the exception of regimes with strong lines, where the observed effect can be tens of percents.

\section{Description of the numerical code}
\label{sec:Code description}

We numerically solve the radiation hydrodynamics equations of a spherically symmetric flow. Following \citet{sapir_numeric_2014}, we allow the photon distribution to deviate from thermal equilibrium with a multi-group treatment, and handle radiative transfer under the diffusion approximation. For the matter equation of state (EOS), we assume an ideal Hydrogen gas in LTE, including ionization as dictated by the Saha equation. The MG emission/absorption and diffusion opacities (effective opacities for each photon energy group), are based on our tables with a solar mix-like composition.

This section is structured as follows. The equations of the numerical scheme are given in \S~\ref{sec:eqs} and \S~\ref{sec:numerical scheme}, and the initial and boundary conditions are described in \S~\ref{sec:init}. The validation of the numeric code and the convergence of the calculations are described in \S~\ref{sec:validation and numerical convergence}.

\subsection{Radiation-hydrodynamics equations}
\label{sec:eqs}

Our equations employ the diffusion approximation and omit terms of order $(v/c)^2$ for radiative transfer, based on \cite{castor_radiation_2007}.
In Lagrangian coordinates, the velocity $\rm v$ and density $\rho$ evolve in response to the radiation energy density $u_\nu$ per unit frequency $\nu$ and matter energy density $e$, as follows:
\begin{equation}
\frac{dr}{dt}=\rm v\label{eq:matter continuity},
\end{equation}
\begin{equation}
    \rho = \rho_{0} \frac{r_{0}^2}{r^2} \frac{\partial r_0} {\partial r},
\end{equation}
\begin{equation}
\frac{d\rm v}{dt}=-\frac{1}{\rho}\frac{d}{dr}\left( P+q\right)+\frac{1}{c}\int_0^\infty \kappa_* j_\nu d \nu
\label{eq: momentum continuity},
\end{equation}
where $P$ is the plasma pressure, $\kappa_*$ is the freq. dept. diffusion opacity, $j_\nu$ is the photon flux, and the $0$ subscript denotes initial values. Correspondingly, the (Lagrangian, fluid element associated) energy densities evolve according to, 
\begin{equation}
    \frac{du_\nu}{dt}=\left.\frac{\partial u_\nu}{\partial t}\right|_{\rm emis/abs}+\left.\frac{\partial u_\nu}{\partial t}\right|_{\rm compres}+\left.\frac{\partial u_\nu}{\partial t}\right|_{\rm diff},
\label{eq:list dudt terms}
\end{equation}
\begin{equation}
    \frac{de}{dt}= -\int_0^\infty \left.\frac{\partial u_\nu}{\partial t}\right|_{\rm emis/abs} d\nu +\left.\frac{\partial e}{\partial t}\right|_{\rm compres}.
\label{eq:list dedt terms}
\end{equation}
The emission / absorption, compression, and diffusion terms are given by

\begin{equation}
\left.\frac{\partial u_\nu}{\partial t}\right|_{\rm emis/abs}=\rho\kappa_{\rm abs,\nu} c \left[\frac{4\pi B_\nu (T)}{c}-u_\nu\right],
\label{eq:emis abs terms}
\end{equation}

\begin{equation}
    \left.\frac{\partial u_\nu}{\partial t}\right|_{\rm compres}=-\left[\frac{4}{3}u_\nu  -\frac{1}{3}\frac{\partial(\nu u_\nu)}{\partial \nu} \right]\frac{1}{r^{2}}\frac{\partial\left(r^{2} {\rm v}\right)}{\partial r},
\label{eq:rad compression term}
\end{equation}
\begin{equation}
\left.\frac{\partial e}{\partial t}\right|_{\rm compres}=-(e+p+q)\frac{1}{r^{2}}\frac{\partial\left(r^{2} {\rm v}\right)}{\partial r},
\end{equation}
\begin{equation}
\left.\frac{\partial u_\nu}{\partial t}\right|_{\rm diffusion}=-\frac{1}{r^2}\frac{\partial (r^2 j_\nu)}{\partial r},
\label{eq:u diffusion term}
\end{equation}
where the frequency-dependent photon energy flux density $j$ is given self-consistently by
\begin{equation}
\label{eq:flux}
    j_\nu=\frac{-1}{\rho\kappa_*}\left(\frac{c}{3}\frac{\partial u_\nu}{\partial r}+\frac{1}{c}\frac{\partial j_\nu}{\partial t}\right),
\end{equation}
and the Planck energy density per unit frequency is given by
\begin{equation}
    \frac{4\pi B_\nu (T)}{c}=\frac{8\pi}{ c^3}\frac{h\, \nu^3}{e^{h \nu/T}-1}.
\end{equation}

$\kappa_*$ is the diffusion opacity, given by the Rosseland mean of sum of the Thomson scattering opacity $\kappa_{\rm es}$ and the absorption opacity $\kappa_{\rm abs,\nu}$ across the frequency bin (of each photon energy group in the MG calculation), $\kappa_*^{-1}= (\Delta\nu_{\rm bin})^{-1}\int d\nu(\kappa_{\rm es}+\kappa_{\rm abs,\nu})^{-1}$, where $\kappa_{\rm abs,\nu}$ is given by our high-resolution opacity tables. The absorption opacity of each group is given by $\kappa_{\rm abs}= (\Delta\nu_{\rm bin})^{-1}\int d\nu\kappa_{\rm abs,\nu}$. Eq. (\ref{eq:emis abs terms}) is based on the assumption that the emitting plasma is in LTE \citep[see e.g. ][ and further discussion in \S~\ref{sec: Summary}]{pinto_physics_2000}.

In a small fraction of the simulations we also include ineslastic Compton scattering using the Kompaneets equation, as shown below. Inelastic Compton scattering is unimportant during shock-cooling, but can have an important effect on the breakout spectrum, which we include in our tests of the code in \S~\ref{sec:validation and numerical convergence}.

\begin{multline}
    \label{eq: u scattering term}
    \left.\frac{\partial u_\nu}{\partial t}\right|_{\rm scat}=-\rho \kappa_{\rm es} c \frac{\nu}{m_{\rm e} c^2} \frac{\partial}{\partial \nu}
    \left[ T \frac{\partial}{\partial \nu} (\nu u_\nu) \right. \\ +  \left. (h\nu - 4 T) u_\nu + \frac{c^3}{8\pi} \left(\frac{u_\nu}{\nu} \right)^2 \right],
\end{multline}
where $m_{\rm e}$ is the electron mass and $T$ is the matter temperature (in units of energy). The flux is initially started as $j_0=\frac{-1}{\rho\kappa_{\rm es}}\frac{c}{3}\frac{\partial u_0}{\partial r}$. The time derivative term in eq.~(\ref{eq:flux}) becomes important in optically thin regions. In our problem it 
does not have an important effect since the luminosity is determined deep in the ejecta.

Following \citet{sapir_numeric_2014}, we rewrite the radiative compression term in eq. ~(\ref{eq:rad compression term}) as
\begin{equation}
    \left.\frac{\partial u_\nu}{\partial t}\right|_{\rm compres}=-u_\nu \left( 1  -\frac{1}{3}\frac{\partial\log u_\nu}{\partial \log \nu} \right)\frac{1}{r^{2}}\frac{\partial\left(r^{2} {\rm v}\right)}{\partial r}.
\label{eq:log rad compression term}
\end{equation}
We include Hydrogen recombination in the EOS, given by
\begin{equation}
    e=\left(\frac{1}{\gamma-1} \right)n(1+Y)T-(1-Y)n I_H, \quad p=n(1+Y)T, 
\label{eq:EOS Hydrogen}
\end{equation}
where $Y(T)$ is the ionization fraction, $I_H$ is the Rydberg energy, $n$ is the atomic number density, $\gamma=5/3$ is the monotonic adiabatic index. This prescription becomes an ideal gas equation of state when the plasma is fully ionized. $Y$ is solved for iteratively using the Saha equation, assuming the presence of Hydrogen only.

\subsection{Numerical scheme}
\label{sec:numerical scheme}
In our numerical scheme, we solve the continuity equations (eqs.~\ref{eq:matter continuity}-\ref{eq: momentum continuity}) by a standard staggered mesh leap-frog method. Energy evolution in time is solved via operator splitting. The equations are divided into parts, with diffusion (including a flux limiter), radiative processes and compression calculated consecutively as follows.

Frequency dependent diffusion, is solved implicitly using a Newton Raphson (NR) solver. The output, including $u_\nu$ and $j_\nu$, is then fed into the radiative processes (emission, absorption, and scattering), which are solved iteratively using two loops of NR solvers that each solve for energy conservation without updating $j_\nu$. In the inner loop, we solve implicitly for $u_{\nu}$, while keeping $e$ constant, while in the outer loop we solve implicitly for $e$, including matter compression. The inner loop includes several protections from non-physical results in $u_{\nu}$, including an `overshoot' protection to prevent $u_{\nu}$ from crossing $B_{\nu}(T)$ when attempting to equalize $u_{\nu}=B_{\nu}(T)$ (see details in \S~\ref{sec:protections diff and rad trans}). The initial guess for $u_{\nu}$ involves solving radiative transfer explicitly, and if the NR solver fails to find a solution after 30 iterations, the solver also attempts a solution starting from the original $u_{\nu}$ value. The energy evolution step finishes with radiative compression and then $j_\nu$ is updated again via eq. \ref{eq:flux}.

Finally, where needed (e.g. eqs. \ref{eq: momentum continuity}, \ref{eq:emis abs terms}, \ref{eq:flux}, $\kappa_*$, $\kappa_{\rm abs,\nu}$), we extract the temperature and pressure by solving the equation of state, eq. (\ref{eq:EOS Hydrogen}) implicitly. The entire set of equations for the evolution of the matter and radiation energies is solved using a predictor-corrector with opacities updated at every iteration prior to the diffusion step.

\subsubsection{Time steps}

The minimum of the following constraints limits our simulation time step. For grid cells $i$, the usual Courant upper limit is $\Delta t_{\rm c}=f_{\rm c}\text{min}\left\{ \Delta x_{\rm i}/C_{\rm s,i}\right\}$, where $\Delta x$ is the grid spacing and $C_{s}$ is the speed of sound. Diffusion also limits the time step according to $\Delta t_{\rm d}=f_{\rm d}\text{min}\left\{ \frac{u_{\nu}}{\partial j_{\nu}/\partial x}\right\} _{\rm i,\nu}$, where the minimum is taken over cells $i$ and bins centered at frequency $\nu$. The factor $f_{\rm d}$, along with all the similar $f$ factors here, is of order unity and smaller than one, and our results are shown to be insensitive to the exact value. Finally, we limit the time step also by limiting the maximal change due to radiative processes of the total energy density of the radiation/plasma,
\begin{flalign}
    \Delta t_{r}=f_{r}
    \begin{cases}
        \frac{u}{j_{B}+j_{C}} & j_{B}+j_{C}<0,\\
        \frac{e}{j_{B}+j_{C}} & j_{B}+j_{C}>0,
    \end{cases}
\end{flalign}
where $u=\int_{0}^{\infty}u_{\nu} d\nu$ is the bolometric radiation energy density. The effective plasma and Compton scattering emissivity are given by $j_{\rm B}=\rho c\int_{0}^{\infty}\kappa_{\nu}\left(B_{\nu}-u_{\nu}\right) d\nu$ and $j_{\rm C}=4u\rho\kappa_{\rm es}c\left(T-T_{\gamma}\right)/m_{\rm e}c^{2}$, where $\kappa_{\nu}$ is the frequency-dependent absorption opacity, $B_{\nu}$ is the blackbody distribution, and T is the matter temperature. The radiation temperature is defined as in \citet{sapir_numeric_2014} as

\begin{equation}
    T_{\gamma}=\frac{1}{4u}\int_{0}^{\infty}\left[h\nu u_{\nu}+\frac{c^{3}}{8\pi}\left(\frac{u_{\nu}}{\nu}\right)^{2}\right]d\nu
\end{equation}

\subsubsection{`Protections' on Diffusion and Radiative Transfer}
\label{sec:protections diff and rad trans}
The very high opacity, reaching $\kappa_\nu\sim10^6 \, \rm cm^2 \, g^{-1}$ at some frequencies, may lead to numeric problems in the application of the implicit solution with finite time steps. At infinitely small time steps, $u_\nu$ will be kept close to $B_\nu$ at such large opacity regions. However, using finite time steps, the implicit result for $u_\nu$ can at times `overshoot' $B_\nu$ (or proceed in the wrong direction due to strong dependence of $\kappa_\nu$ on temperature). In order to avoid impractically short time steps, we add several limiting `protections' immediately after the inner Newton Raphson solver for $u_{\nu}$. Namely, if the resulting implicit $u_\nu$ lies outside of the range between $u_\nu$ of the previous step and $B_\nu$, we override the result to the nearest of these values \citep[see also,][]{mihalas_foundations_1999,mcclarren_semi-implicit_2008,gasilov_solution_2016}. We also limit the change during the emission/absorption time step to $\left|\Delta u_{\rm abs,nu}/u_\nu\right|<f_{\rm abs}$, where $f_{\rm abs}$ is in the range $\{0.1,0.5\}$. Its value does not affect our results. Though the latter constraint affects the relative rates of physical processes, absorption still proceeds quickly relative to the other processes in this scheme.

We also add a flux limiter to the simulations. The P1 diffusion approximation doesn't require one in principle. However, at certain frequencies, strong absorption and the strong sensitivity of $\kappa_{\nu}$ to temperature, can lead to a situation where the photon energy density in a particular frequency group and cell depletes abruptly -for example to match $B_{\nu}(T)$-. The change may occur faster than the time in which diffusion can respond given finite time resolution, and thus may lead to non-physical flow. Namely, either flux would flow in the wrong direction, or $\left|j\right|>u_{\rm cell}c$, where $u_{\rm cell}$ is the energy density in the cell from which the flux $j$ is exiting, and $c$ is the speed of light. In lieu of adding another time constraint, we insert a flux limiter as follows,
\begin{equation}
    j_\nu\to j_{\nu,\rm FL}=j\:[1+\left(\left|j_\nu\right|/u_{\rm cell}c\right)^{m}]^{-1/m}
\end{equation}
where $m$ is a positive integer constant.
The derivative for $j$ -defined in between cells- with respect to u, in either one of the adjacent cells, is written as
\begin{equation}
    \frac{\partial j_{\nu, \rm FL}}{\partial u}=\frac{\alpha}{\left[1+\left|\gamma\right|^{m}\right]^{1/m}}-\frac{\gamma^{m-1}}{\left[1+\left|\gamma\right|^{m}\right]^{1/m+1}}
\left[\frac{j_\nu}{\left|j_\nu\right|}\alpha-\left|\gamma\right|c\frac{\partial u}{\partial u_{\rm cell}}\right],
\end{equation}
where $\alpha\equiv\partial j_\nu/\partial u$ and $\gamma\equiv j_\nu/u_{\rm cell}c$. This derivative is used in the Newton Raphson solver during the diffusion step.

We test the flux limiter on the gray diffusion runs with various values of $m$, finding a deviation of less than a percent from the non-flux limited runs when $m=8$, which we choose in our simulations.

\begin{figure}
    \centering
    \includegraphics[width=\columnwidth]{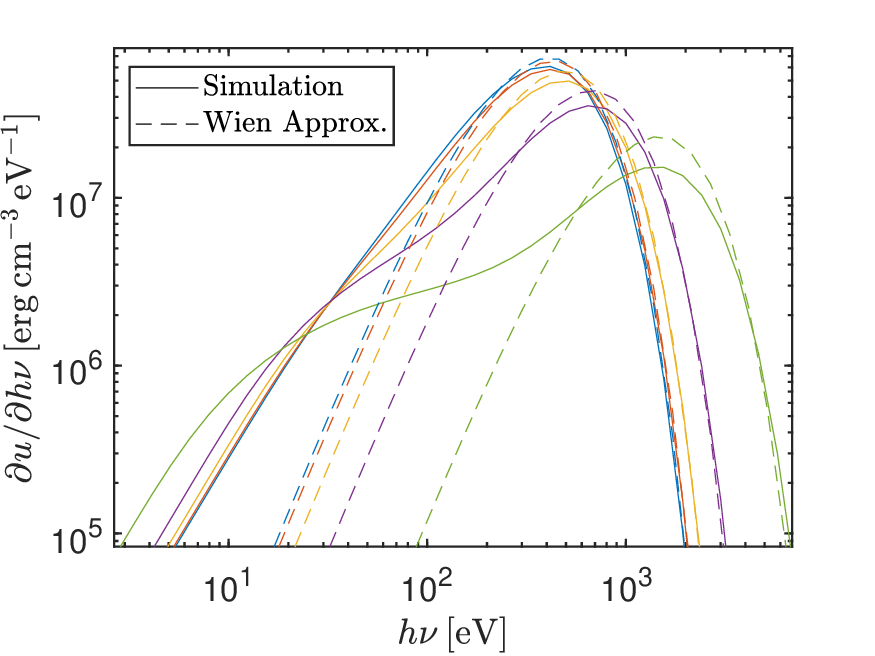}
    \includegraphics[width=\columnwidth]{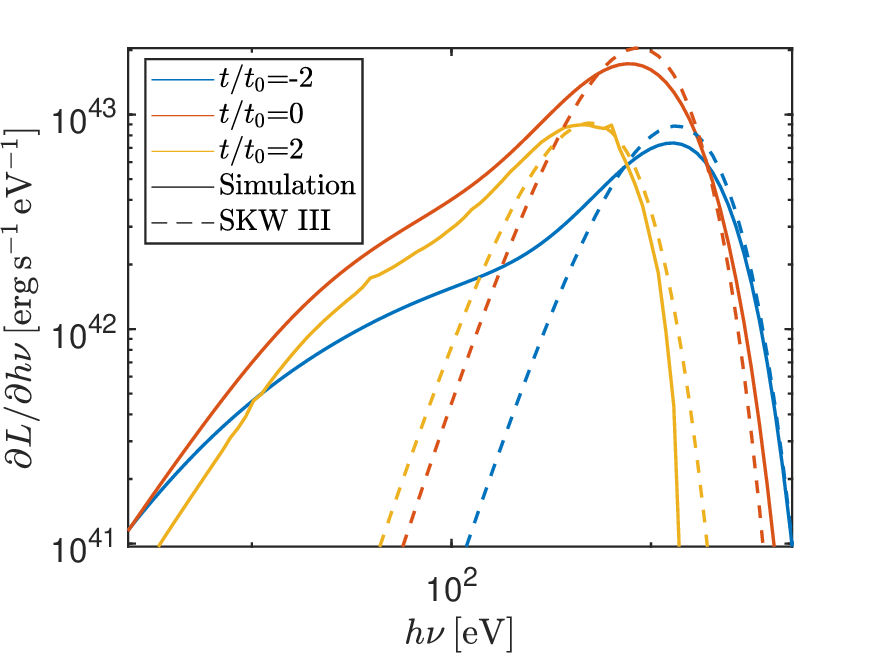}
    \caption{Tests of our multigroup code. Top: an example of a radiation mediated shock (RMS) in a uniform medium, with photon energy density $\partial u/\partial h\nu$, shown at several locations across the shock. This is compared with an analytic Wien profile, defined by the analytic bolometric energy density $u$, and the local photon number density $n_\gamma$, extracted from simulation. The shock is of velocity $\beta$=10\%, in a pure Hydrogen medium, with initial proton number density of $n_{\rm p}=10^{15} \, \rm g \, cm^{-3}$ . Bottom: luminosity $\partial L/\partial h\nu$, at shock breakout (over the course of several shock-crossing times $t_0$) with a solar composition and a free-free opacity only. The simulation is compared at breakout time to the table results of \citetalias{sapir_non-relativistic_2013}, where we extract breakout velocity, density, and breakout time based on a fit of the bolometric luminosity to \citetalias{sapir_non-relativistic_2011} (as is done in \citetalias{morag_shock_2023}), without any additional fitting. The progenitor radius is $R=10^{13}$ cm, core and envelope masses are $M_{\rm env}=M_{\rm core}=10M_\odot$, and the explosion energy is $E=10^{52} \, \rm erg \, cm^{-3}$.}
    \label{fig:RMS example}
\end{figure}

\subsection{Initial and boundary conditions}
\label{sec:init}

Our numeric calculation involves a succession of three simulations, with each simulation starting later in physical time using a snapshot of the hydrodynamic variables as described by its predecessor. Each successive simulation also contains increasing physical complexity; i.e. first a hydrodynamic-only calculation, then a gray diffusion calculation, and finally a MG simulation. This way, later time stages of interest include all the relevant processes, while allowing the computations to be performed in practical time. All simulations are proceeded through to the latest times for comparison.

Following \citetalias{sapir_uv/optical_2017} and \citetalias{morag_shock_2023}, we begin with a simplified progenitor structure, comprising of a uniform density core surrounded by a polytropic envelope at hydrostatic equilibrium. We start a hydrodynamic only simulation where we inject a high thermal energy density in the innermost cells of the core and capture the resulting shock using artificial viscosity. Then a radiative diffusion-hydrodynamics gray simulation is started between 24 and 8 shock crossing times prior to shock breakout, with the exact start time have negligible effect in this range. Both of the simulations are identical to the ones in \citetalias{morag_shock_2023} (and described there in detail), with the important exception that in the gray simulation we increase the initial cell resolution towards the stellar edge\footnote{At the edge we use a grid spacing $\Delta r_{\rm grid} \propto (R-r)$, down to a predetermined scattering optical depth of $\tau_{\rm es} \lesssim 10^{-3}$. The additional edge resolution has no effect on the results of the gray simulation and allows us to improve edge resolution in the subsequent MG calculation.}.

Then we begin a multi-group diffusion simulation, as described in \S~\ref{sec:numerical scheme}. The simulation is started between 20 shock crossing times prior to breakout and up to 2 R/c times after breakout, with the exact time at which the simulation is started having negligible effect on the later shock-cooling emission. Typical resolutions for each of the respective stages listed above are 4000-8000, 1600-3200, and 200-1600 cells (MG runs that are started after shock breakout typically have 50-200 cells), with 32-256 photon groups in the MG phases. All calculations are continued until at least after the recombination time $t_{0.7 \rm eV}$.

Multigroup simulations that are started prior to breakout have a similar initial grid to the gray diffusion simulations. Namely the grid changes smoothly, with modest resolution in the interior, highest resolution at the starting location of the shock, and steadily decreasing resolution outwards \citepalias[keeping cell count constant across the RMS - see ][]{sapir_uv/optical_2017}, before approaching a constant for $\tau\lesssim c/v_{\rm bo}$, and finally increasing resolution at the stellar edge, with $\Delta r \sim (R-r)$, down to at least a scattering optical depth of $\tau_{\rm es}\sim 10^{-2}$. MG simulations that were started after shock breakout have a simpler initial cell grid, spaced logarithmically in $\tau_{\rm es}$, with the same stellar edge resolution. All MG simulations have photon frequency bins that are constant in time and are spaced logarithmically.

For all simulations we assume a static reflective boundary in the inner surface, and a free boundary at the outer surface that for the diffusion simulations accelerates as $\partial_t v_{\rm b} = j_{\rm b}\kappa / c$, where the subscript b denotes boundary values. The boundary flux is given by  $j_{\rm b}=f_{\rm edd}cu_{\rm b}$, where $f_{\rm edd}=0.3-0.5$ is the Eddington factor. The results are insensitive to the exact value of $f_{\rm edd}$ since the flux is determined deep within the plasma, at $\tau\sim c/v\gg1$.

\subsection{Code validation and numerical convergence}
\label{sec:validation and numerical convergence}

In \citetalias{morag_shock_2023}, we validated our numerical hydrodynamical-only code against the analytic planar stellar breakout solutions of \citet{gandelman_shock_1956} and \citet{sakurai_problem_1960}, and our gray diffusion code against the analytic planar "Sakurai-Weaver" Anzats solutions of \citetalias{sapir_non-relativistic_2011}. We also reproduced the bolometric breakout flux expected from a planar stellar breakout, as also described in \citetalias{sapir_non-relativistic_2011}. In \citet{sapir_numeric_2014}, an earlier version of the multi-group code that we use here, underwent several test problems involving radiative diffusion, emission/absorption, and inelastic Compton scattering.

We perform here two additional tests of the multigroup code; the problem of a steady planar radiation mediated shock, and the breakout spectrum in a hydrogen dominated stellar envelope, both including inelastic Compton scattering and only free-free absorption opacity.

We calculate the structure of the steady planar radiation mediated shock at two representative velocities, $\beta={\rm v}/c=1\%$ and 10\%, i.e. spanning breakout velocities in our parameter range. We find for both cases that density $\rho$, velocity $\rm v$, and bolometric photon energy density $u$, converge to $3\%$, $0.5\%$, and $\sim1\%$ relative to the analytic result \citep[see][and \citetalias{sapir_non-relativistic_2011}]{weaver_structure_1976}. For the $\beta=1\%$ simulation, the photons are in LTE, and $u_\nu$ is in excellent agreement with a Planck distribution matching total bolometric luminosity $u$. For $\beta=10\%$, the photons carrying most of the energy are in Compton equilibrium. While an exact analytical solution does not exist for the spectrum in this case, we find good agreement between the numeric results for the frequency of the peak of the radiation energy density behind the shock and an analytic estimate using a Wien distribution\footnote{The Wien distribution is determined by two parameters, the photon energy density $u$, which is determined analytically, and the number density of photons $n_\gamma$, which we extract from the simulation.}, see figure \ref{fig:RMS example}. At lower frequencies, the energy density transitions to a thermal distribution due to large free-free opacity, producing a visible deviation from the Comptonized spectrum, as is expected.

Next, we compare our results for envelope breakout with the approximate table values from \citetalias{sapir_non-relativistic_2011,sapir_non-relativistic_2013}, which again assume a fully Comptonized Wien spectrum. We find reasonable (10's of \%) agreement in peak temperature and luminosity (fig. \ref{fig:RMS example}), which is somewhat remarkable, since the temperature and luminosity profiles in these tables are a function of only the breakout parameters ($R,\rho_{\rm bo},\beta_{\rm bo}$ - see \S~\ref{sec: Physics of RMS and SCE}). For the comparison we extract these parameters from the simulation without performing additional fitting for the SED. There is again a noticeable deviation in the low energy tail due to thermalization. \citet{sapir_numeric_2014} performed MG diffusion calculations of the planar envelope breakout phase, and obtained similar results to ours. They find 10's of \% agreement with \citetalias{sapir_non-relativistic_2011,sapir_non-relativistic_2013} in peak temperature and luminosity, as well as a thermalized low-frequency tail of similar shape.

In \citetalias{morag_shock_2023}, we showed convergence of the hydrodynamic and gray simulations. Our MG calculations are also converged with respect to spatial resolution. Doubling the spatial resolution produces at most a few percent change (and often less than $1\%$) in $L_\nu \equiv \partial L/\partial \nu$. We also verify that we are converged with respect to the resolution of the outermost cell, finding that $L_\nu$ varies by less than $1\%$ when the minimum scattering optical depth varies between $\tau_{\rm es} \sim 10^{-2}$ and $10^{-3}$, in agreement with the conclusions of \citet{tominaga_shock_2011}.
We note that our frequency resolution is high enough that our SED is insensitive to the number of photon frequency groups, but is coarse relative to the atomic line scale, which we discuss at length in \S~\ref{sec: finite frequency resolution}.

As described in \citetalias{morag_shock_2023}, we include in both the gray and MG diffusion simulations, a non-radiating plasma component coupled with artificial viscosity $q$. This addition helps stabilize against numerical instabilities associated with the density inversion that occurs at the outer edge of the ejecta.

\section{Our composite opacity table}
\label{sec:kappa}

Calculating the frequency dependent opacity requires employing several approximations and assumptions. Primary challenges involve solving the many-electron Schrödinger equation and estimating microplasma interactions between species. The assumption that all degrees of freedom are in thermal equilibrium is often, though not always, employed. Due to these approximations, there are large uncertainties in the opacity. For example, a factor of 2 discrepancy in the Rosseland mean exists between TOPS and OP \citep{iglesias_updated_1996} in our regime of interest (as shown in \citetalias{morag_shock_2023}).

We built our own frequency-dependent opacity table, containing free-free, bound-free, and bound-bound components, and assuming local thermal equilibrium. The code that produces these tables is now available to the community on github \cite{morag_frequency_2023} and produces tables at arbitrary density, temperature, and frequency resolution (we use $\Delta\nu/\nu\sim 10^{-5}-10^{-6}$ in practice). The opacity code does not include "expansion opacity" effects.

In \citetalias{morag_shock_2023} we used the Rosseland mean opacity from the high-resolution tables to provide a formula for blackbody emission. Here we bin the tables in frequency to produce our multigroup opacities. For the absorption term ($\kappa_{\rm abs,\nu}$ in eq. \ref{eq:emis abs terms}) we use the average opacity across each bin, while for the diffusion term ($\kappa_*$ in eq.~\ref{eq:u diffusion term}) we take the Rosseland mean across the bin. In this work, we include 10 important atoms up to iron: H, He, C, N, O, Ne, Mg, Si, S, Fe (though the opacity code can handle an arbitrary mixture). We show later that the resulting supernova lightcurves are insensitive to the exact composition in the case of Hydrogen dominated envelopes.

As discussed in \citetalias{morag_shock_2023}, the frequency-dependent TOPS opacities are verified against experiments at temperatures and densities exceeding tens of eV and $10^{-6} \text{g cm}^{-3}$ respectively \citep{colgan_light_2015,colgan_new_2016,colgan_new_2018}. The Kurucz database of atomic transitions was calibrated against measured line frequencies and oscillator strengths \citep{kurucz_atomic_1995}, but is incomplete for highly ionized species at high, $T>10$~eV, temperature. We separately run simulations using both our opacity table and using TOPS.

This section is written as follows. In \S~\ref{sec: opacity components}, we describe the construction of the opacity, and in \S~\ref{sec: tests and results of the opacity} we describe tests and results of the table.

\begin{figure}
    \centering
    \includegraphics[width=\columnwidth]{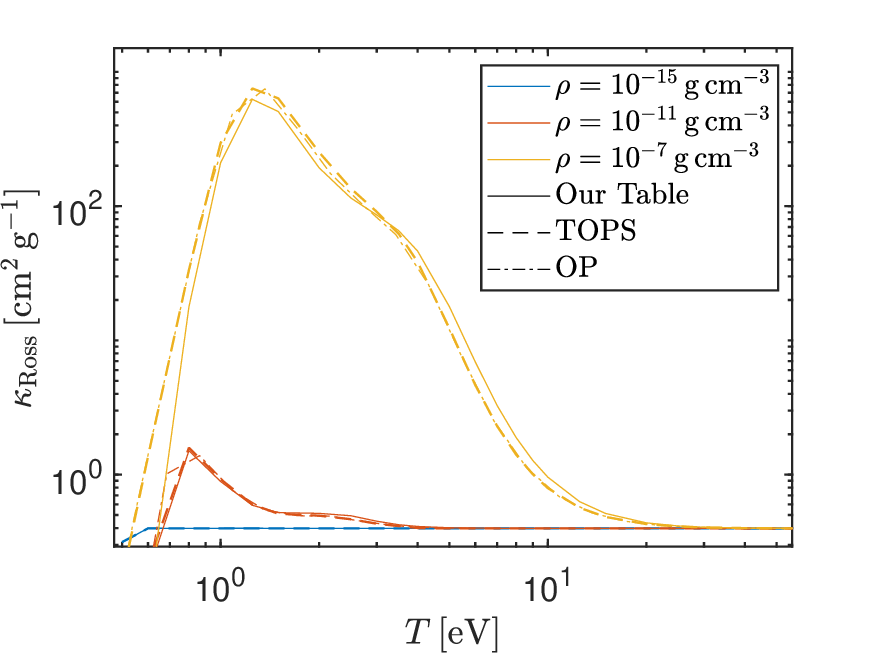}
    \includegraphics[width=\columnwidth]{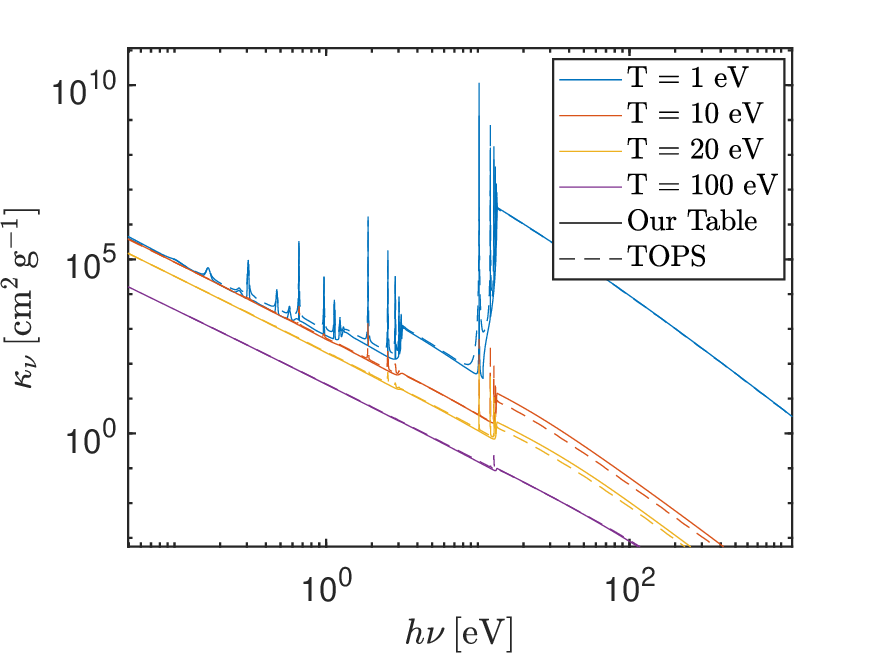}
    \caption{A comparison of our opacity table code results against TOPS and OP results for pure Hydrogen, showing excellent agreement. At top, a comparison of the Rosseland mean at different densities and temperatures. At bottom, example frequency dependent opacities at a density of $\rho=10^{7} \, \rm  g \, cm^3$.}
    \label{fig:H ours vs TOPS}
\end{figure}

\begin{figure}
    \centering
    \includegraphics[width=\columnwidth]{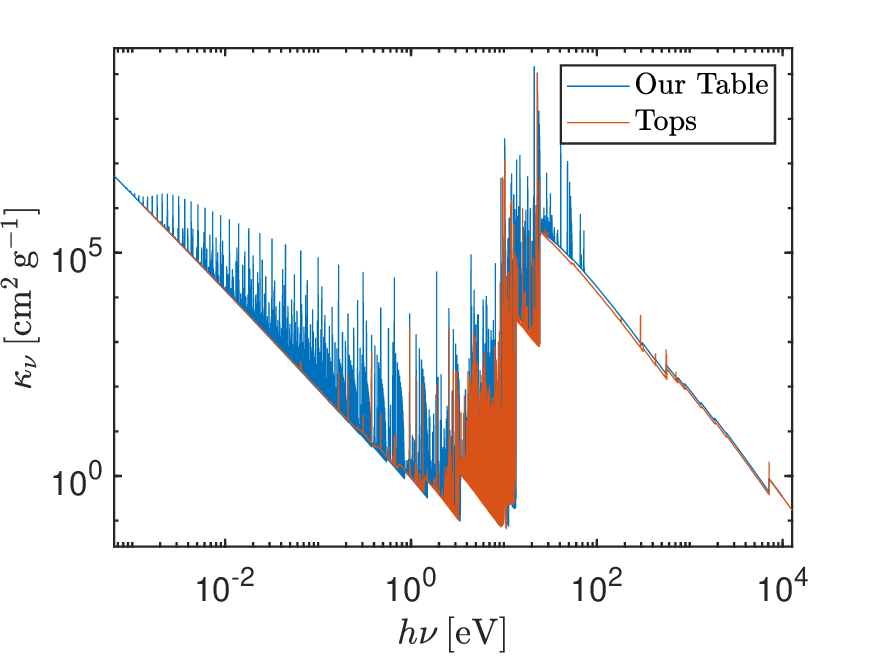}
    \caption{An example of our frequency dependent opacity table for a solar mix at a density of $\rho=10^{-11} \, \rm g \, cm^{-3}$ and temperature $T=1$ eV, compared with TOPS, showing reasonable agreement. The observed discrepancy in the lower-frequency atomic lines is most likely a result of the lower resolution in TOPS, where such lines cannot be seen. Since these lines are very narrow, the effect on the MG simulations is small.}
    \label{fig:Solar Mix ours vs TOPS}
\end{figure}

\subsection{Opacity - Construction}
\label{sec: opacity components}
We solve for the ionization and excitation population numbers self-consistently with the Saha equation, and using electron levels provided by the NIST database\footnote{\url{https://www.nist.gov/pml/atomic-spectra-database}}. For the free-free components and Hydrogen-like bound-free and bound-bound components, we use the equations provided in \citet{rybicki_radiative_1979}. We verify our free-free calculations against TOPS, finding 7\% agreement or better.

For line transitions in Hydrogen-like ions with charge $Z$, the decay rate is given by $A(Z)\sim\nu^{2}f=Z^{4}A(Z=1)$, where $A(Z=1)$ is the decay rate for the equivalent Hydrogen line, and  $\nu\sim Z^{2}$ is the frequency of the atomic transition.  Non Hydrogen-like bound-free absorption is based on the table in \citet{verner_atomic_1996}\footnote{This table only includes photoionization from electrons in the ground-state. To calculate the electron ground occupation levels, we include all states up to 0.3 eV above the ground state to account for hyperfine splitting.}. In general, the results of the table agree w/TOPS up to a factor of 2, with the exception of a pure Fe mix we tested, where we observe an order of magnitude difference. Bound-bound  oscillator strengths, degeneracies and lower energy levels are taken from Kurucz CD 23\footnote{Kurucz line lists are expanded semi-regularly, representing ongoing progress. Our simulation results are insensitive to the choice of line list, including CD 23 and the most up-to-date file from Oct. 8th 2017. In cases of mismatch between the reported lower and upper state energy gap and the transition frequency, we use the lower state and the transition frequency. Missing decay rates are estimated analytically where available. For nearly all lines, the natural line width is smaller than the thermal width, which in turn is smaller than our grid resolution.}.

Our line sampling method varies with line width relative to the grid resolution $\Delta\nu_{\rm grid}$. Lines with full width half max (FWHM) that are much thinner than the grid spacing, $\Delta \nu$, are sampled as a single grid point, with magnitude proportional to  $\sim (\Delta \nu)^{-1}$ to conserve total flux. Broad lines are sampled as a Voight function $\phi_\nu$, including native line width and thermal broadening. Intermediate width lines are sampled from the frequency derivative of $\int \phi_\nu d\nu$ such that oscillator strength is conserved. We find that we are insensitive to the exact choice of cutoff between each of the sampling regimes in the ranges 1/30<FWHM/$\Delta\nu_{\rm grid}$<1/3 and 1<FWHM/$\Delta\nu_{\, \rm grid}<10$, respectively. We therefore choose the cutoffs at FWHM/$\Delta\nu_{\, \rm grid}=1/10$ \footnote{To avoid cases where the Lorentz wings can have an affect, we also require the FWHM of the Lorentzian (not including thermal broadening) to be $3\times10^3$ times thinner than $\delta \nu_{\rm grid}$ to sample as a delta function. We also don't use the delta function for any of the Hydrogen lines.} and 3.

For quicker calculation in the latter two cases, the broad Voight wings are interpolated at 100 times coarser resolution than the frequency grid, and combined at the end of the calculation\footnote{We are insensitive to the location of the cutoff between fine and coarse sampling resolution. Nominally, we place the cutoff 3 FWHM's away from the center.}. Finally, ions and electronic states with electron occupation fractions below $10^{-14}$ are removed (we tested that this omission has no effect).

We also include Red-wing continuum suppression \citep[following TOPS -][]{colgan_new_2018}, and 
the Hummer-Mihalas factor that suppresses higher electron states \citep{hummer_equation_1988}.
Our opacity code allows computation of other ingredients that were not added in this work as they were shown to have negligible influence on the multigroup simulations in our range of parameters, including the Dappen Anderson Mihalas factor \citep{dappen_statistical_1987}, and electron collisional broadening. These latter processes are described in the documentation for the opacity.

\subsection{Tests of the Opacity Calculation and Results}
\label{sec: tests and results of the opacity}
We test our frequency-dependent opacity code against TOPS and OP for the simplest case of a pure Hydrogen mixture, finding good agreement ($ \lesssim 15\%$) in both Rosseland mean and frequency-dependent opacity (see figs. \ref{fig:H ours vs TOPS} and~\ref{fig:Solar Mix ours vs TOPS}). As a further sanity check, we also compare our bound-bound opacity for a pure Fe mix with an independent code from \citet{waxman_constraints_2018}, finding excellent agreement (not shown). We find that our simulations are converged with respect to the underlying opacity table, as tested in more than 5 separate parameter choices spanning progenitor radius and explosion energies for 128 photon groups. We observe $<2\%$ difference in the SED when varying between $\Delta\nu/\nu \sim 10^{-5}-10^{-6}$ base grid, and $<1\%$ when changing the number of \{$R\equiv\rho/T^3,T$\} grid points from \{16,66\} to \{30,120\}.

\section{Deviations from Blackbody - Calibrated Analytic Model}
\label{sec:analytic formula}

Here we provide an analytic description of the deviations of the emitted spectrum from blackbody. 
Our approximation formulae are derived piecewise, based on the strength of the absorption opacity, $\kappa_{\rm abs,\nu}$, relative to the scattering opacity, $\kappa_{\rm es}$. At frequencies where $\kappa_{\rm abs,\nu}>\kappa_{\rm es}$, the emitted flux may be approximated as a blackbody with a frequency-dependent thermal depth (surface of last absorption) $r_{\rm col,\nu}$, and corresponding frequency-dependent color temperature $T_{\rm col,\nu}=T(r_{\rm col,\nu})$, given by
\begin{equation}
    L_{\nu,\rm BB}=4\pi r_{\rm col,\nu}^2 B_{\nu}(T_{\rm col,\nu}).
    \label{eq: freq dept blackbody Tcol rcol prescription}
\end{equation}
At frequencies where the absorption opacity is smaller than the scattering opacity, $\kappa_{\rm abs,\nu}<\kappa_{\rm es}$, we base our approximation on the flux $f_\nu$ emitted by a semi-infinite planar slab of temperature $T$ in the two-stream approximation \citep{rybicki_radiative_1979},
\begin{equation}
 f_{\nu}=\frac{4\pi}{\sqrt{3}}\frac{\sqrt{\epsilon_{\nu}}}{1+\sqrt{\epsilon_{\nu}}}B_{\nu}(T),
    \label{eq: fnu semi infinite slab}
\end{equation}
where
\begin{equation}
    \epsilon_\nu=\kappa_{\rm abs,\nu}/(\kappa_{\rm abs,\nu}+\kappa_{\rm es}).
    \label{eq: epsilon def}
\end{equation}
We therefore suggest 
\begin{equation}
 L_{\nu,\epsilon}=\frac{\left(4\pi\right)^{2}}{\sqrt{3}}r_{col,\nu}^{2}\frac{\sqrt{\epsilon_{\nu}}}{1+\sqrt{\epsilon_{\nu}}}B_{\nu}(T_{col,\nu}),
    \label{eq: epsilon prescription}
\end{equation}
as an approximation for the escaping spectral luminosity in this regime \citep[see also, ][]{dessart_distance_2005}.

We first derive an expression describing the emission at the regime of relatively low absorption opacity, $\kappa_{\rm abs,\nu}<\kappa_{\rm es}$, which occurs primarily at intermediate frequencies near and below the Planck peak (e.g. $h\nu \sim 1-3$~eV in fig.~\ref{fig:Solar Mix ours vs TOPS}). Absorption in this regime is dominated by free-free transition with a small bound-free contribution. Neglecting the bound-free contribution we approximate eq.~(\ref{eq: Thermal Depth Integral k_abs_nu}), that defines the frequency-dependent thermal depth, as \citepalias[see also,][]{shussman_type_2016}
\begin{equation} 
\int_{r_{\rm col,\nu}}^{\infty}\rho\sqrt{3\kappa_{\rm ff,\nu}\kappa_{es}}dr' =1.
    \label{eq: Thermal depth integral kff limit}
\end{equation}
Here we have neglected $\kappa_{\rm abs,\nu}$ with respect to $\kappa_{\rm es}$ and used
\begin{multline}
    \kappa_{\rm abs,\nu}\to \kappa_{\rm ff,\nu}=\\
    4.13\times10^{-31}g_{\rm ff}\rho T^{-1/2}(h\nu)^{-3}\left(1-\exp\left(-h\nu/T\right)\right) \, \rm cm^2 \, g^{-1},
\end{multline}
where the density $\rho$ is in cgs, temperature $T$ is in ergs. We approximate the gaunt factor $g_{\rm ff}=\frac{\sqrt{3}}{\pi}K_{0}(h\nu/T)\sim0.717\left(h\nu/T\right)^{-0.27}$, with $K_{0}$ being the zeroth modified Bessel function of the second kind.

Solving eq.~(\ref{eq: Thermal depth integral kff limit}) using the analytic \citetalias{rabinak_early_2011}/\citetalias{sapir_uv/optical_2017} spherical phase density profiles (eqs. \ref{eq:rho_rt_vsstar} - \ref{eq:T_rt_bo}) we obtain the radius, temperature and opacity at the thermal depth,
\begin{equation}
    r_{\rm col,\nu} = 1.29 \times 10^{14} R_{13}^{-0.01}
    (f_{\rho}M_0)^{0.09} {\rm v_{\rm s*,8.5}^{0.78}}
    \kappa_{0.34}^{0.03} t_{\rm d}^{0.80} \nu_{eV}^{-0.08} \, \rm cm,
\end{equation}
\begin{equation}
    T_{\rm col,\nu} = 2.13 \, R_{13}^{0.28}
    (f_{\rho}M_0)^{-0.02} {\rm v_{s*,8.5}^{0.17}} \nu_{\rm eV}^{0.25} \kappa_{0.34}^{-0.08}
    t_{\rm d}^{-0.42} \, \rm eV,
\end{equation}
\begin{equation}
    \kappa_{\rm ff,\nu} =
    0.02 \, (f_\rho M_0)^{-0.05}
    R_{13}^{-0.23} {\rm v_{\rm s*,8.5}^{-0.66}}
    \kappa_{0.34}^{-0.29}
    t_{\rm d}^{-0.19}
    \nu_{\rm eV}^{-1.66} \,
    \rm cm^2 \, g^{-1}.
\end{equation}
We find that modifying the expression for $r_{\rm col,\nu}$ to
\begin{equation}
    r_{\rm col,\nu} = R + 1.29 \times 10^{14} R_{13}^{-0.01}
    (f_{\rho}M_0)^{0.09} {\rm v_{\rm s*,8.5}^{0.78}}
    \kappa_{0.34}^{0.03} t_{\rm d}^{0.80} \nu_{eV}^{-0.08} \, \rm cm
\end{equation}
while keeping the expressions for $T_{\rm col,\nu}$ and $\kappa_{\rm ff,\nu}$ unchanged provides a good description of the spectrum also at the planar phase (and at the transition from planar to spherical evolution),  
In break notation we have:
\begin{equation}
    r_{\rm col,\nu} = R + 2.18 \times 10^{13} L_{\rm br,42.5}^{0.48} T_{\rm br,5}^{-1.97}
    \kappa_{0.34}^{-0.07} \tilde{t}^{0.80} \nu_{\rm eV}^{-0.08} \, \rm cm,
    \label{eq: r_col_nu_br_notation}
\end{equation}
\begin{equation}
    T_{\rm col,\nu} = 5.47 \, L_{\rm br,42.5}^{0.05} T_{\rm br,5}^{0.92}
    \kappa_{0.34}^{0.22}
    \tilde{t}^{-0.42} \nu_{\rm eV}^{0.25} \, \rm eV,
    \label{eq: T_col_nu_br_notation}
\end{equation}
\begin{equation}
    \kappa_{\rm ff} = 0.03 \, L_{\rm br,42.5}^{-0.37} T_{\rm br,5}^{0.56} 
    \kappa_{0.34}^{-0.47}
    \tilde{t}^{-0.19}
    \nu_{\rm eV}^{-1.66} \, \rm cm^2 \, g^{-1},
    \label{eq: k_col_nu_br_notation}
\end{equation}
where $\tilde{t}=t/t_{\rm br}$ and $R$ is given in terms of the break parameters by
\begin{equation}
    R = 2.40\times10^{13} \, t_{\rm br,3}^{-0.1} \,L_{br,42.5}^{0.55} \,T_{br,5}^{-2.21} \,\text{cm}.
    \label{eq: R_of_br_params}
\end{equation}
The thermal depth values can then be inserted into eq. (\ref{eq: epsilon prescription}) with $\epsilon_\nu=\kappa_{\rm ff,\nu}/(\kappa_{\rm ff,\nu}+\kappa_{\rm es})$ in order to describe the emission in the low absorption frequency range.

For frequency regions with strong absorption, where $\kappa_{\rm abs,\nu}<\kappa_{\rm es}$, we return to eq. (\ref{eq: freq dept blackbody Tcol rcol prescription}). The thermal depth at these frequencies is located at the outer edge of the ejecta, where the density decreases sharply and the temperature, determined by the free-streaming photons, is nearly uniform. We therefore approximate $r_{\rm col,\nu}\approx const. (\nu)$ and $T_{\rm col,\nu}\approx const. (\nu)$ for these frequencies, and describe the emission as a gray blackbody $L_{BB}$, eq. (\ref{eq:L_nu_BB_formula}). At low frequencies where the free-free opacity dominates, we find numerically that the luminosity is well approximated by $L_{BB}(0.85 \, T_{\rm col})$. Meanwhile, at frequencies near and above the Planck peak, where atomic transitions dominate, we use both the simulations and a separate analytic estimate (see \S~\ref{sec: finite frequency resolution}) to improve upon the approximate $L_{\rm BB} (0.74 \, T_{\rm col})$ description of the UV suppression of \citetalias{morag_shock_2023}, replacing the suppression factor $0.74$ with a function of $(R,t)$ lying in the range $[0.6,1]$.

The combined freq-dept formula is thus
\begin{equation}
\label{eq:Lnu epsilon final}
L_{\nu} = \begin{cases}
\left[L_{\rm BB} (0.85 \, T_{\rm col})^{-m} + L_{\nu,\epsilon}^{-m}\right]^{-1/m} & h\nu<3.5 T_{\rm col} \\
1.2 \times L_{ \rm BB}(0.85 R_{13}^{0.13} t_d^{-0.13} \times T_{\rm col}) & h\nu>3.5 T_{\rm col},
\end{cases}
\end{equation}
where $m=5$, and $L_{\nu,\epsilon}$ is again given by eqs. (\ref{eq: epsilon prescription}) and (\ref{eq: epsilon def}) with the choice $\kappa_{\rm abs,nu} \to \kappa_{\rm ff,\nu}$. The 1.2 factor accounts for modest UV excess we observe in our results at the planck peak due to the presence of strong lines. The frequency slope in the Raleigh-Jeans regime is similar, but slightly lower than the blackbody value $L\sim\nu^2$.

Eq. (\ref{eq:Lnu epsilon final}) can be further simplified to be given in terms of only $L$ and $T_{\rm col}$, with a minor decrease in the approximation's accuracy, 
\begin{multline}
   L_{\nu} = \\
   \begin{cases} \frac{\pi}{\sigma}\frac{L}{T_{\rm col}^4} \left[\, \left(\frac{B_{\nu}(0.85 T_{\rm col})}{(0.85)^4} \right)^{-m} + \right.\\
        \left. \left( \frac{8}{\sqrt{3}} x^{-0.155} T_{\rm col,5}^{-0.1} \frac{\sqrt{\epsilon_{\rm a}}}{1+\sqrt{\epsilon_{\rm a}}} B_\nu(1.63 \, x^{0.247} T_{\rm col}) \right)^{-m} \right]^{-1/m} & h\nu<3.5 T_{\rm col} \\
        \quad & \quad \\
        1.2\times L_{\rm BB}(1.11 L_{42.5}^{0.03} T_{5}^{0.18} \times T_{\rm col}) & h\nu>3.5T_{\rm col},
   \end{cases}
   \label{eq:Lnu epsilon simplified}
\end{multline}
where $x=h\nu/T_{\rm col}$, $T_{\rm col} = 5 \, T_{\rm col,5} \, \rm eV$, and $\epsilon_{\rm a} = 0.0055 \, x^{-1.664} T_{\rm col,5}^{-1.0996}$. $L$ and $T_{\rm col}$ are given by eqs. (\ref{eq:L_trans}) and (\ref{eq:T_trans}).

\section{Numeric results}
\label{sec:numeric_res}

\begin{figure*}
    \centering
    \includegraphics[width = \textwidth]{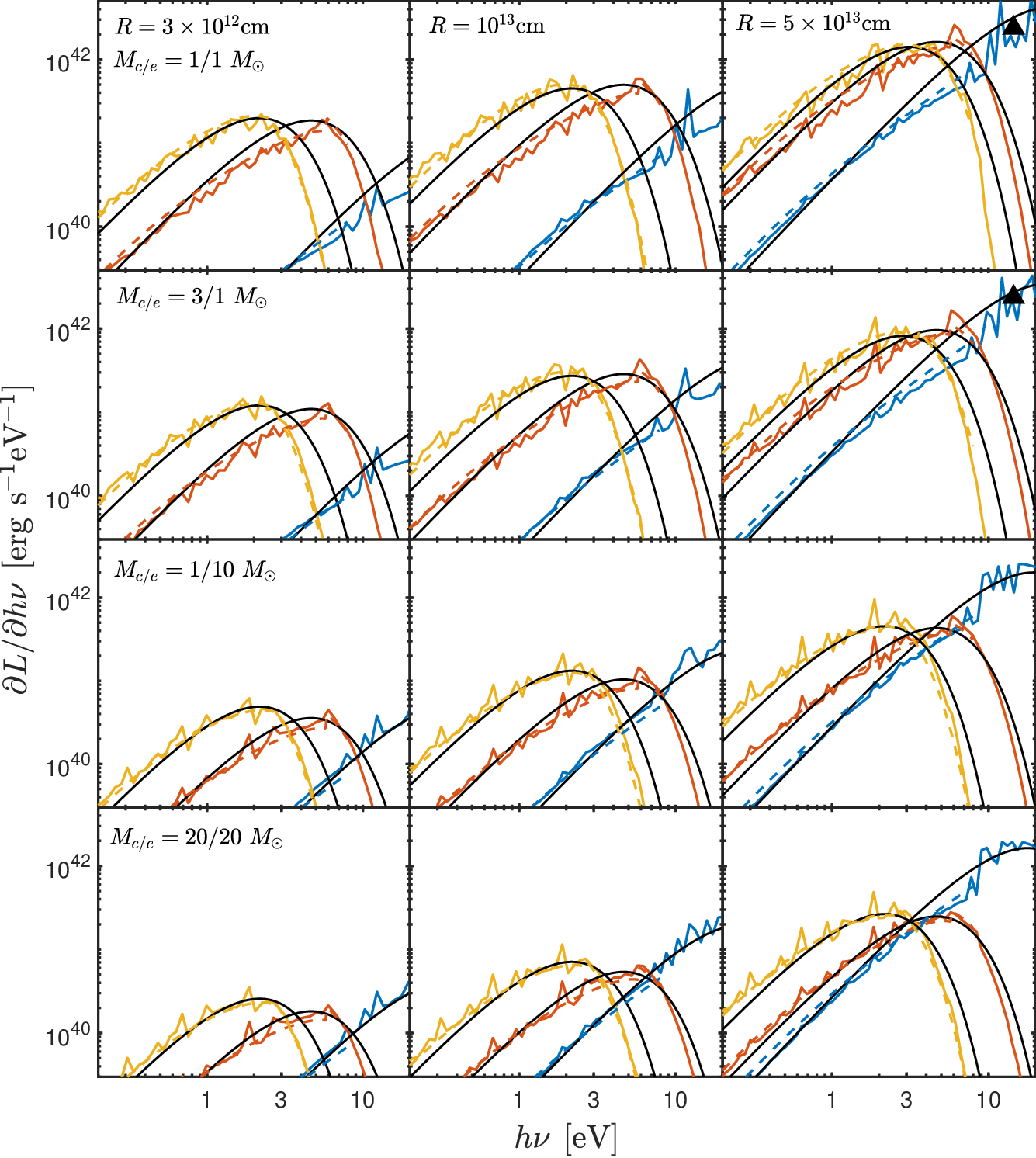}
    \caption{Observed luminosity $\partial L/\partial h\nu$, shown for a range of progenitor radii and core/envelope mass combinations, comparing our analytic formulas with multigroup simulation results. Colored solid lines represent simulations. Our analytic formula from \citetalias{morag_shock_2023} (eqs. \ref{eq:L_trans}, \ref{eq:T_trans}) is in black, and our modified frequency-dependent formula (eq. \ref{eq:Lnu epsilon final}) is in dashed colored lines, showing excellent agreement with the simulations. The blue / red / yellow color denotes results at times $3R/c$, $t_{1.5 \rm eV}$, $t_{0.7 \rm eV}$. In plots with with black triangles, the time $t_{\rm 0.7 \, eV}$ (in yellow) is replaced by the transparency validity time $t_{\rm tr}/a$, since the latter occurs prior to recombination. The formula is based on breakout parameters, ($\rho_{\rm bo},\beta_{\rm bo},R$), extracted from simulation without additional fitting involved.}
    \label{fig:Lnu_tiles_lowR}
\end{figure*}

\begin{figure*}
    \centering
    \includegraphics[width = \textwidth]{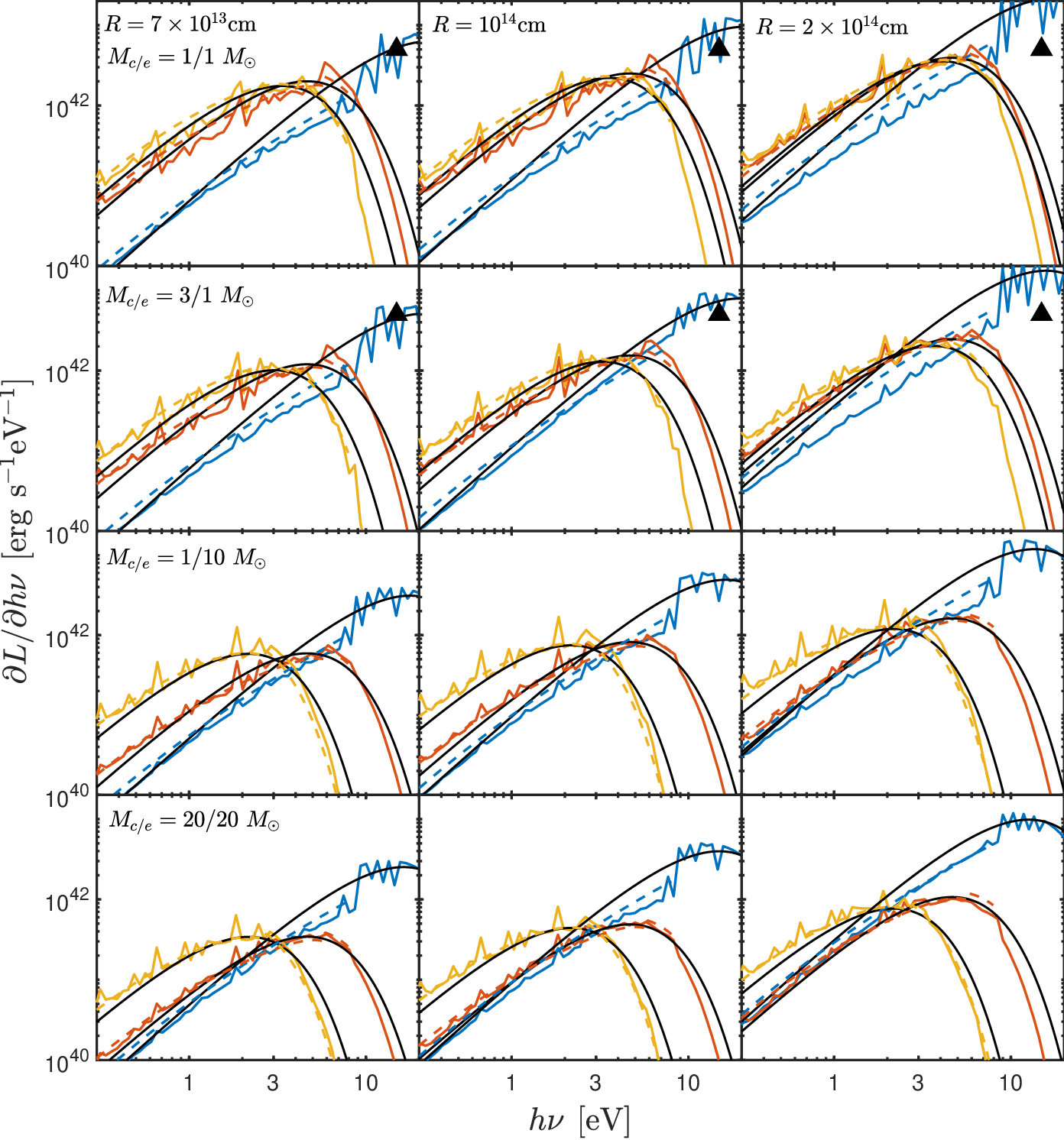}
    \caption{Same as fig. \ref{fig:Lnu_tiles_lowR}, but showing larger radii.}
    \label{fig:Lnu_tiles_hiR}
\end{figure*}

In figs. \ref{fig:Lnu_tiles_lowR} and \ref{fig:Lnu_tiles_hiR} we plot shock-cooling numeric results from over a dozen multigroup simulations. Deviations from our gray blackbody formula (eqs.~\ref{eq:L_trans}-\ref{eq:L_nu_BB_formula}) are small. In the Raleigh Jeans regime, the SED slope is slightly shallower than $L \sim \nu^2$. As frequency increases in this regime, $L_\nu$ passes from 10's of percents above the Planck distribution to 10's of percents or more below it. Near the Planck peak at $h\nu \sim 3.5$ eV, the SED can be 10's of \% in excess of the prediction of our blackbody formula (50\% in extreme cases, further discussed in \S~\ref{sec: finite frequency resolution}). In the Wien tail, the flux is suppressed due to line absorption.

The SED obtained numerically is also compared to our frequency dependent formula (eq. \ref{eq:Lnu epsilon final}) yielding good agreement, with an RMS error of $\Delta L_\nu/L_\nu\lesssim20\%$ for $h\nu<3 T_{\rm col}$ (and several 10's of \% or more in the Wien tail, reflecting a very small inaccuracy in the radiation temperature). The frequency-dependent formula is generally closer to the simulation results than the gray formula prediction throughout. Both formulas shown in the figure use breakout parameters ($\rho_{\rm bo}$,\, $\beta_{\rm bo},\, R$) that are derived by comparing the breakout bolometric luminosity to \citetalias{sapir_non-relativistic_2011} (as was done in \citetalias{morag_shock_2023}), without additional fitting.

We find in many simulations that the ratio of bolometric luminosity between the multigroup and gray simulations is approximately given by $(\kappa_{\rm es}/\kappa_{\rm Ross})$, where the Rosseland mean opacity $\kappa_{\rm Ross}$ is evaluated at the diffusion depth, where $L$ is determined (recall that the gray simulation only include Thomson opacity). This factor is less important during the planar phase, but generally reduces $L$ in the multigroup simulation by $10's$ of \% during the spherical phase, primarily due to Wien suppression. 

\begin{figure}
    \label{fig:us vs TOPS}
    \label{fig:Lnu_metallicity}

    \includegraphics[width = \columnwidth]{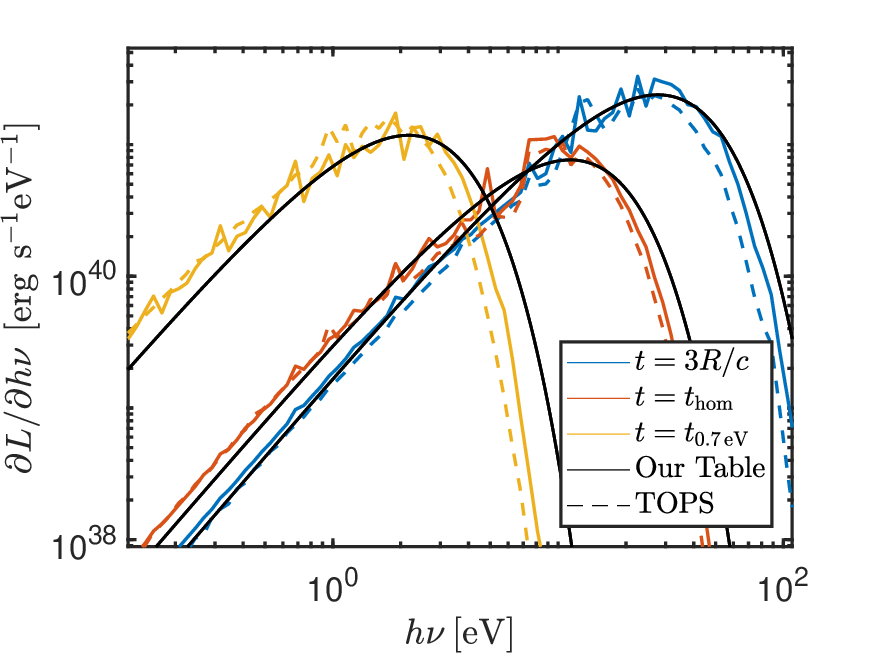}
    \includegraphics[width=\columnwidth]{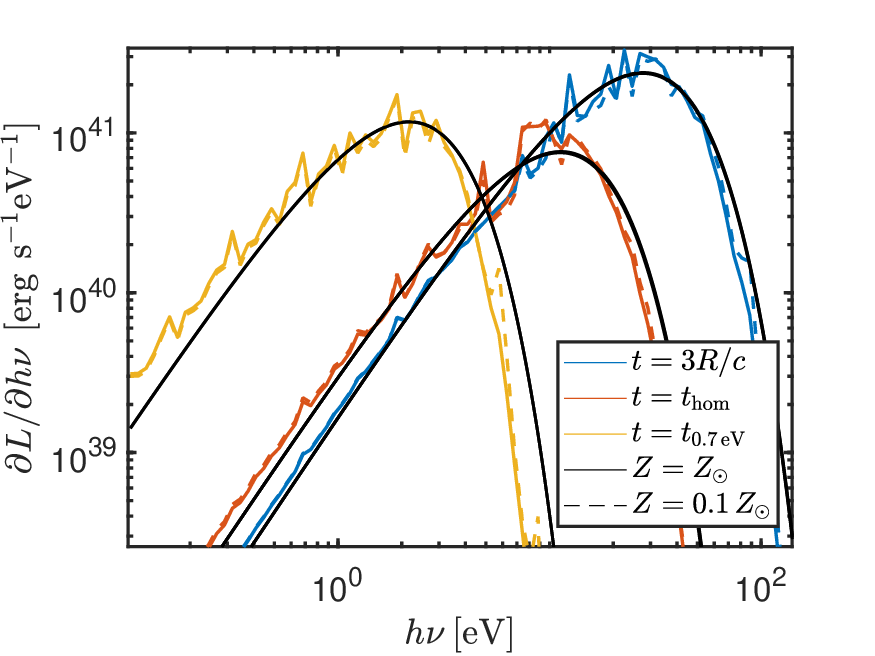}
    
    \caption{Observed luminosity, same as fig. \ref{fig:Lnu_tiles_lowR}, showing example checks of the sensitivity to opacity changes. At top, comparison using our table and TOPS, showing $\sim10\%$ difference in $T_{\rm col}$ at late times, and otherwise minimal effect. At bottom, the effect of changing metallicity, exhibiting negligible effect. Both panels are showing simulations with the parameters $R=10^{13}$ cm, $E = 10^{51}$ erg, and core and envelope masses of $M_{\rm core}=M_{\rm env}=10 M_\odot$.}
    \label{fig:Lnu_TOPS_vs_ours}
\end{figure}

We test the sensitivity of our results to the choice of opacity table using simulations that span the progenitor radius and explosion energy parameter range. When we use TOPS table instead of our own, we find that near recombination time, $T_{\rm col}$ in the presence of the TOPS-derived opacity can be  lower by up to 10's of \% (usually <5\% - see example in fig.~\ref{fig:us vs TOPS}). This result is in general agreement with our gray analysis in \citetalias{morag_shock_2023}, where  we concluded that for $T_{\rm col}<4$ eV, it is preferable to use our opacity table due the presence of lab-confirmed lines. For higher temperatures, early during shock-cooling, we find negligible (few percent) SED difference between the two opacity tables, since most of the observable frequencies ($h\nu \lesssim10$ eV) are in the Rayleigh Jeans regime, and are less affected by the presence of lines.

We also test the sensitivity to composition in our simulations, finding a negligible variation when metallicity varies from $Z=0.1$ solar to solar metallicity (see fig. \ref{fig:Lnu_metallicity}), in agreement with \citetalias{morag_shock_2023} results.  We conclude that the SED in Hydrogen dominated envelopes is insensitive to metallicity. We note, however, that for near zero metalicity, up to a factor of 2 differences may be obtained at UV frequencies.

\section{Expansion opacity, finite frequency resolution, deviations from LTE}

\subsection{Expansion opacity and finite frequency resolution}
\label{sec: finite frequency resolution}

Our numeric calculation method did not include "expansion opacity" effects and assumed that the opacity, $\kappa_\nu$, does not vary significantly across the extent $\Delta r$ of a single spatial resolution element. The large velocity gradients in the flow we are considering, combined with the presence of strong absorption lines, may lead to significant variations of $\kappa_\nu$ across $\Delta r$ due to the space dependent Doppler shift. This effect may have a significant impact on the calculation of photon transport. For the diffusion calculation, we use a Rosseland average of the opacity, $\kappa_{R,i}=1/\overline{\kappa^{-1}}$ where the average is over the frequency band $\nu_i$-$\nu_{i+1}$, implying a photon mean free path of $l_R=1/(\kappa_{R,i}\rho)$. The Rosseland mean is dominated by the frequencies at which the opacity is low, i.e. where lines are absent, and hence the presence of strong lines does not affect $l_R$ significantly. In the presence of a large velocity gradient, the photon mean free path may become significantly shorter- as the photon propagates through the varying velocity of fluid, the frequencies of the lines shift and the photon will be absorbed when it reaches a position where the shifted frequency of a strong line coincides with the photon's frequency \citep{friend_stellar_1983,Eastman_Spectrum_1993,castor_radiation_2007,rabinak_early_2011}. In other words, the effective Doppler broadening "closes" the low $\kappa_\nu$ "windows" in frequency space, that allow a low value of $\kappa_R$ and large $l_R$. The absorption mean free path is given in this case by
\begin{equation}\label{eq:l_exp}
  l_{\rm exp}\approx\frac{c}{\rm v}\frac{\Delta\nu}{\nu}r,
\end{equation}
where $\Delta\nu$ is the frequency separation between strong lines, and we have used $\partial v/\partial r=v/r$ as appropriate for the homologous expansion phase. A line is considered "strong" if it leads to $\tau>1$ when integrated over the photon propagation path taking into account the Doppler shift- for line opacity $\kappa_\nu=\kappa_l\nu_l\delta(\nu-\nu_l)$, $\tau_\nu=\int dr \rho\kappa'=\kappa_l\rho ct(\nu_l/\nu)$ (where $\kappa'$ is the Doppler shifted opacity). Thus, $\Delta\nu$ is the frequency distance between lines with 
\begin{equation}\label{eq:strong_l}
    \kappa_l\rho ct>1.
\end{equation}

$l_{\rm exp}$ can be used to define an effective "expansion opacity", taking into account the finite velocity difference across $\Delta r$ due to the expansion of the plasma. The ratio between the Rosseland and expansion mean free paths is 
\begin{equation}\label{eq:lRlexp}
  \frac{l_{\rm exp}}{l_R}=(\kappa_R\rho r)\frac{c}{\rm v}\frac{\Delta\nu}{\nu}
  =(\kappa_R\rho r)\frac{c}{\rm v}\left(\nu\frac{dN}{d\nu}\right)^{-1}\approx 10\tau_R\frac{c}{\rm \rm v}\left(\nu\frac{dN}{d\nu}\right)^{-1},
\end{equation}
where $dN/d\nu$ is the (strong) line density and we have used $\rho\propto r^{-10}$ (as appropriate for the shock cooling expanding plasma) yielding $10\tau_R=\kappa_R\rho r$. 

\begin{figure}
    \centering
    \includegraphics[width=\columnwidth]{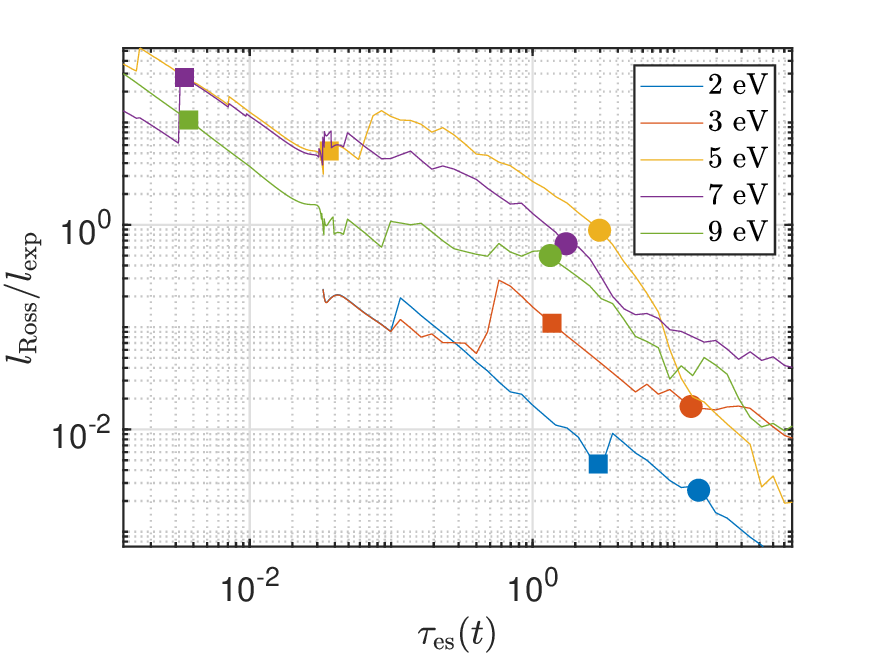}
    \includegraphics[width=\columnwidth]{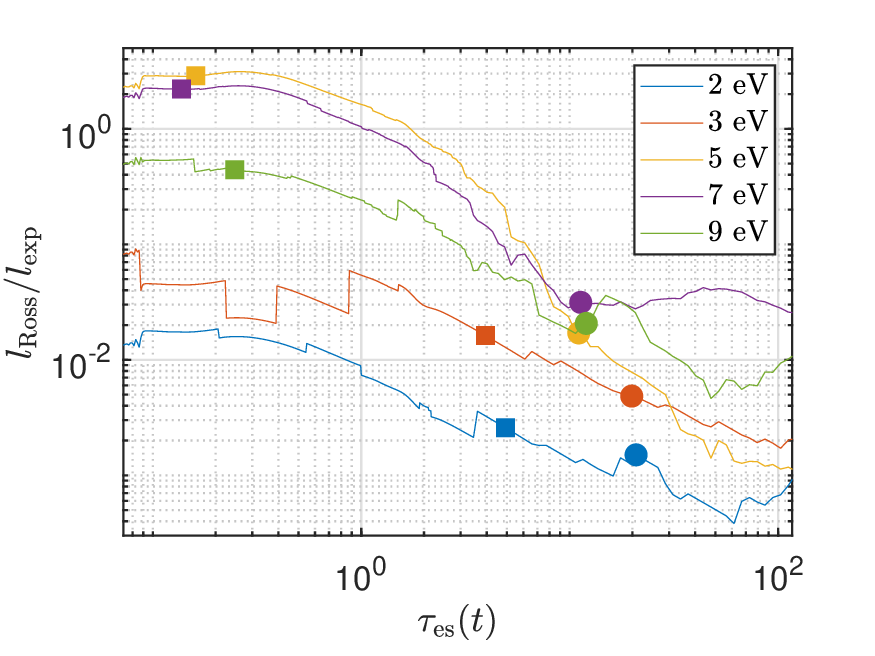}
    \caption{Rosseland mean-free-path relative to expansion mean-free-path as a function of scattering optical depth $\tau_{\rm es}$. Different photon frequency groups are shown in different colors, with the $\bullet$ and $\blacksquare$ symbols indicating the location of the diffusion depth and thermal depth (estimated without expansion opacity) respectively. Results are shown at the time of recombination $t_{0.7 \, \rm eV}$, when the effect of expansion opacity is strongest, for two different progenitor radii, $R=3\times10^{12}$~cm (top) and $10^{14}$~cm (bottom), with explosion energy $E=10^{51}$ erg, and core and envelope masses $M_{\rm core}=M_{\rm env}=10M_\odot$.}
    \label{fig:lRoss_lDopp}
\end{figure}

In the region where the escaping flux is determined, $\tau_R\approx c/\rm v$, we have $l_{\rm exp}/l_R\approx10(c/{\rm v})^2(\Delta\nu/\nu)\approx 10^{3.5}(\Delta\nu/\nu)$. We therefore expect $l_{\rm exp}/l_R\gg1$ in this region, and thus only a negligible effect of the "expansion opacity" on the escaping flux. This is demonstrated to be the case in Fig.~\ref{fig:lRoss_lDopp}, showing $l_R/l_{\rm exp}$ for various photon energies near recombination time as a function of $\tau_{\rm es}$, the electron scattering optical depth. As demonstrated in the figure, the suppression of the mean free path due to the "expansion opacity" effect at this time may affect the spectrum at photon energies $>5$~eV, since $l_{\rm exp}/l_R<1$ may be obtained at the thermalization depth of such high energy photons. We arrive at the same conclusion from examination of simulations spanning different progenitor radii and explosion energies.

\begin{figure}
    \centering
    \includegraphics[width=\columnwidth]{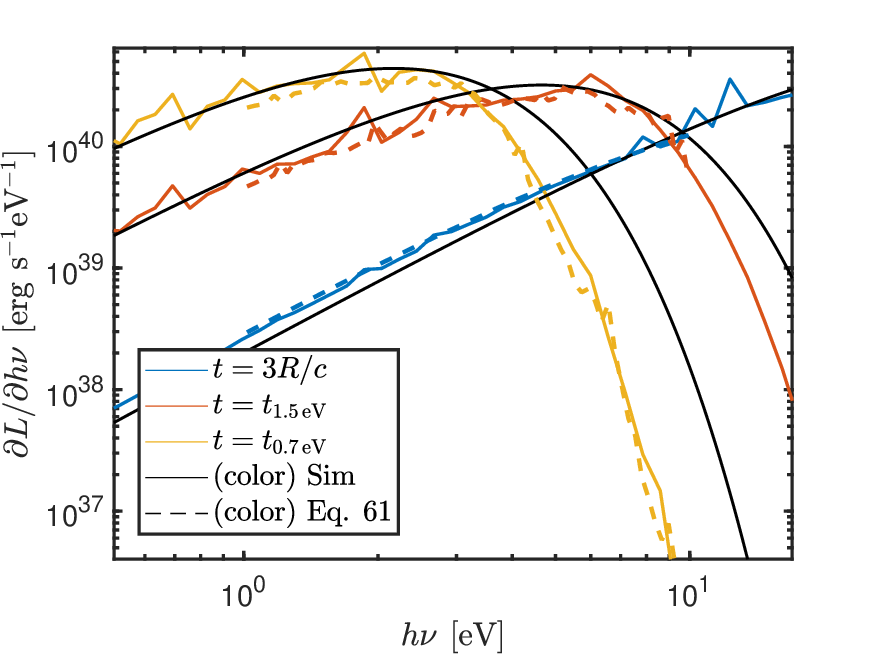}
    \includegraphics[width=\columnwidth]{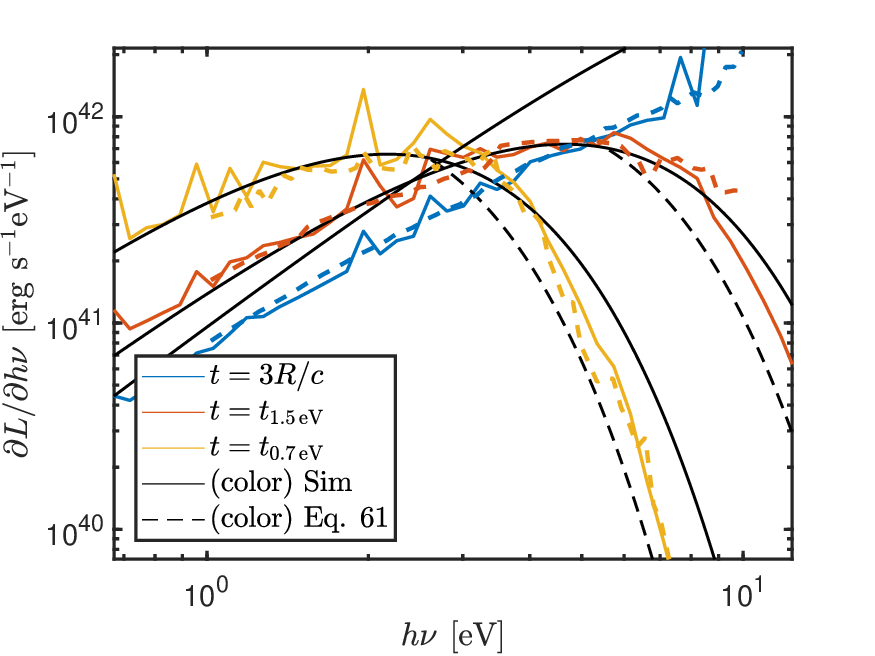}
    \caption{A comparison of the flux obtained in the numeric simulations, that do not include the effects of expansion opacity, with the flux obtained using the analytic approximations of eqs.~(\ref{eq: freq dept blackbody Tcol rcol prescription}) and~(\ref{eq: epsilon prescription}) with $r_{\rm col,\nu}$ (eq.~\ref{eq: Thermal Depth Integral k_abs_nu}) calculated with the density and temperature profiles obtained in the simulation, but with a high resolution, $\Delta\nu/\nu\sim10^{-5}$ opacity table and taking into account the Doppler shifts of the lines 
    (The gray approximation, eqs. (\ref{eq:L_trans})-(\ref{eq:L_nu_BB_formula}) shown in black). The high-resolution spectrum has been averaged over frequency bins, without affecting the SED.
    Top (bottom): progenitor radius $R = 3\times10^{12}$ cm ($R=10^{14}$ cm), explosion energy $E=10^{51}$ erg, and core and envelope masses $M_{\rm env}=M_{\rm core}=10 M_{\odot}$.}
    \label{fig:R03_R10_Lnu_vs_thrmdpth_Doppler}
\end{figure}

In order to test the possible impact of the expansion opacity on the high energy spectrum, we compare (see fig.~\ref{fig:R03_R10_Lnu_vs_thrmdpth_Doppler}) the flux obtained in the numeric simulation, which does not include the effects of expansion opacity, with the flux obtained using the analytic approximations of eqs.~(\ref{eq: freq dept blackbody Tcol rcol prescription}) and~(\ref{eq: epsilon prescription}) with $r_{\rm col,\nu}$ (eq.~\ref{eq: Thermal Depth Integral k_abs_nu}) calculated with the density and temperature profiles obtained in the simulation. In the latter we use a high resolution, $\Delta\nu/\nu\sim10^{-5}$ opacity table and take into account the Doppler shifts of the lines\footnote{$T_{\rm col,\nu}$, and $\epsilon_{col,\nu}$ are determined at $r=r_{\rm col,\nu}$, and we approximate $\epsilon_{\rm col,\nu}= \tau_{\rm abs,\nu} / (\tau_{\rm abs,\nu} + \tau_{\rm es})$, where the optical depths again include Doppler shifting and are evaluated up to the thermal depth, $(\tau_{*,\nu}=1)$.
}.
The SED is given by
\begin{equation}
   L_{\nu,\rm Dopp} = 
   \begin{cases}
        \rm Eq. \, (\ref{eq: freq dept blackbody Tcol rcol prescription}) & \tau_{\rm es} ( \tau_{*,\nu}=1) \leq f_{\rm cut} \\
        \rm Eq. \, (\ref{eq: epsilon prescription}) & \tau_{\rm es} ( \tau_{*,\nu}=1) > f_{\rm cut},
   \end{cases} 
   \label{eq: Doppler Integral}
\end{equation}
where $f_{\rm cut}$ determines the scattering optical depth $\tau_{\rm es}$ at the thermal depth below which we neglect the effect of scattering. The results are not sensitive to the choice of $f_{\rm cut}$ in the range 0.3-3. We show results for $f_{\rm cut}=1$.

The analytic approximation, with $r_{\rm col}$ calculated using a high resolution frequency-grid and including expansion opacity effects,
reproduces well the spectrum obtained in the simulations.
The modifications of $r_{\rm col}$, and the implied flux modification in the analytic approximation, due to the inclusion of the Doppler shifts of lines using a high resolution frequency grid, yield therefore an estimate of the magnitude of this effect. The good agreement between the simulations results and the analytic estimate obtained as described above implies that the simulation, with its coarse frequency resolution, reasonably estimates the effect of the expansion opacity with an accuracy of typically 10\%.

The comparison of the analytic and numeric results described above, reveals several frequency bins at intermediate photon energies, 5-8~eV, in which the flux obtained in the numeric simulation exceeds that which is obtained by the analytic estimate by a factor of a few. This is due to the presence of strong and relatively isolated lines, which lead to a large average absorption opacity $\kappa_{\rm abs,i}$ of the photon bin, which in turn leads to a blackbody photon spectrum across the frequency bin with temperature corresponding to the plasma temperature. Since the radiation energy density at intermediate photon energies is below the plasma temperature at radii where the optical depth is small (see \S~\ref{sec:NLTE} and fig. \ref{fig:NLTE u_nu B_nu example}), the flux obtained in bin $i$ is significantly larger than the flux obtained at other frequencies. This result is an artifact of the numeric calculation using finite frequency resolution- the emitted flux would match the blackbody spectrum (of the plasma temperature) only across the (very) narrow line width, hardly affecting the total flux in the finite frequency range of bin $i$. In order to remove this effect, we have identified and removed a few such isolated lines from the opacity table used in the simulations. 
The 5 removed lines are listed in table \ref{Table:Removed Lines List}. Much of the remaining discrepancy between the simulations and the analytic estimate is likely due to the fact that the flux obtained in the numeric simulation exceeds that of the analytic results due to an overestimate of the effect of lines.

\begin{table}
\centering
\begin{tabular}{ |p{1.6cm}||p{2.8cm}| }
 \hline
 Species &  Transition Energy [eV] \\
 \hline
 O III & $4.967$ \\
 C III & $5.398$ \\
 N IV & $7.215$ \\
 C IV & $7.995$ \\
 C IV & $8.009$ \\
 \hline
\end{tabular}
\caption{List of lines removed from our finalized simulations (see \S~\ref{sec: finite frequency resolution}).}
\label{Table:Removed Lines List}
\end{table}

\subsection{Deviation from LTE Ionization and Excitation}
\label{sec:NLTE}

Deviations of plasma excitation and ionization states from LTE (`NLTE' effects) may become pronounced, and have been studied by many groups, at late times, in particular during the nebular phase \citep{baron_non-local_1996,utrobin_strong_2005,dessart_time-dependent_2008,lisakov_study_2017}. NLTE effects at the very early times that we are considering, prior to H recombination, are less widely studied and are expected to be small \citep[see, e.g.][]{blinnikov_radiation_2000,baron_preliminary_2000,dessart_time-dependent_2008}\footnote{\citet{takeda_non-lte_1990,takeda_formation_1991} predicts order of magnitude deviations of atomic electron populations from LTE at $T_{\rm ph}\sim1$~eV. We note, however, that these calculations are performed for densities, which are orders of magnitude smaller than those relevant for the shock-cooling problem that we are considering. The Takeda 1990 photospheric radius, $10^{14}$~cm, implies a photospheric density of $\sim10^{-13}$~g/cc (for the steep $1/r^{10}$ density profile expected for the expanding stellar envelope). This is nearly 3 orders of magnitude higher than the density used in the Takeda 1990 calculation.}. We briefly explain below why the LTE assumption is reasonable for modeling the SED in the first hours and days.

\begin{figure}
    \centering
    \includegraphics[width=\columnwidth]{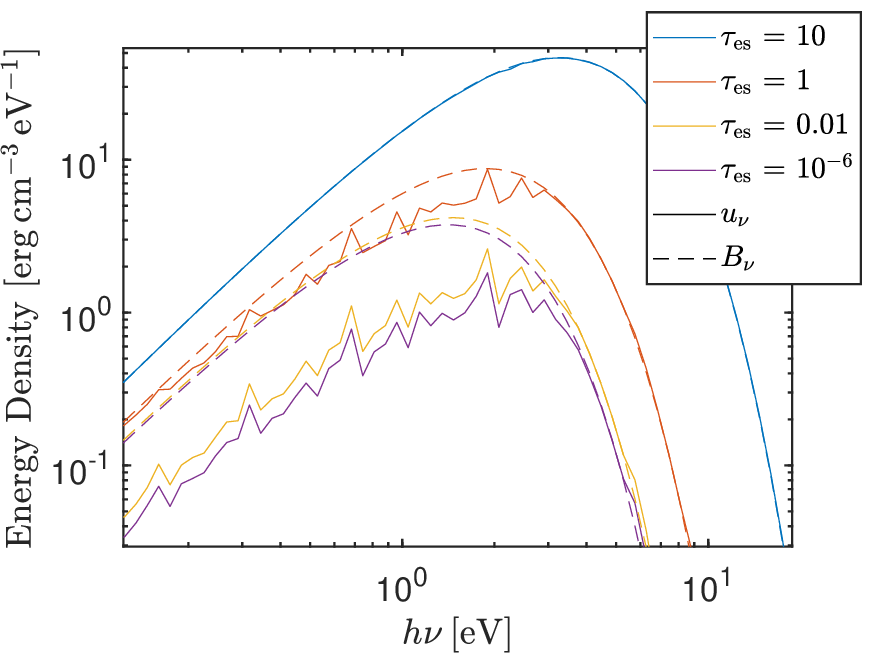}
    \caption{
    Example of the photon energy distribution $u_\nu$ compared to the equivalent Planck distribution at the local plasma temperature.
    Photons match the local temperature deep in the ejecta and for frequencies $h\nu>3$ eV. Elsewhere, they are lower than the local Planck distribution because they are dominated by photons that have arrived from deeper in the ejecta and have been suppressed by scattering. The example shown is of an explosion with progenitor radius $R=10^{13}$ cm, core and envelope masses of $M_{\rm env}=M_{\rm core}=10M_\odot$, and explosion energy $E=10^{51}$ erg, shown at recombination time $t_{0.7 \, \rm eV}$.}
    \label{fig:NLTE u_nu B_nu example}
\end{figure}

The density of the plasma emitting the radiation during the hours to days of shock breakout and cooling is relatively high, and the radiation is close to thermal equilibrium with the plasma - see fig.~\ref{fig:NLTE u_nu B_nu example}. This situation is quite different from that prevailing later, on weeks time scale, where the density is low and the radiation is far from thermal equilibrium, causing the ionization fraction and excitation level distribution to deviate largely from a Boltzmann distribution. 

\begin{figure}
    \centering
    \includegraphics[width=\columnwidth]{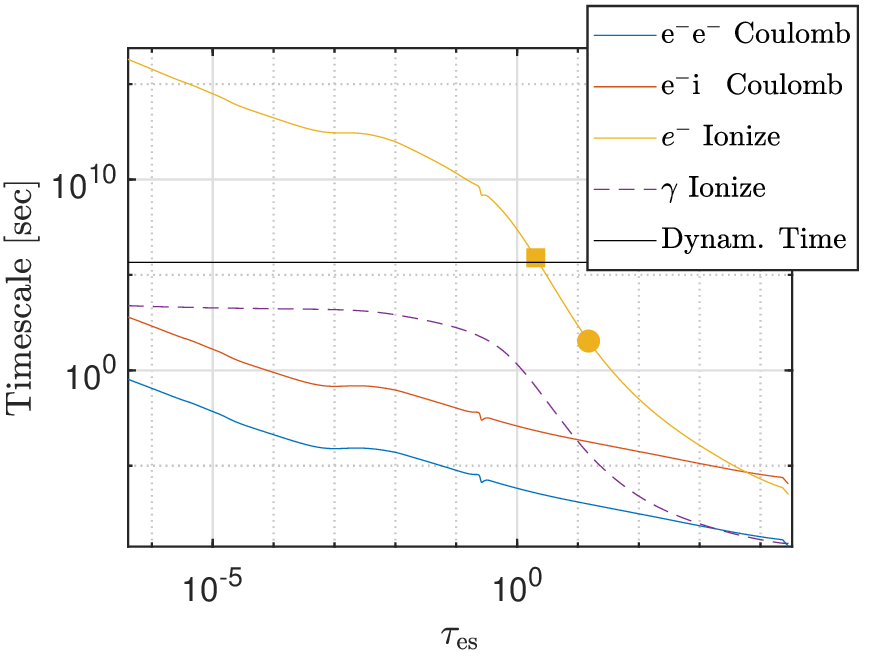}
    \includegraphics[width=\columnwidth]{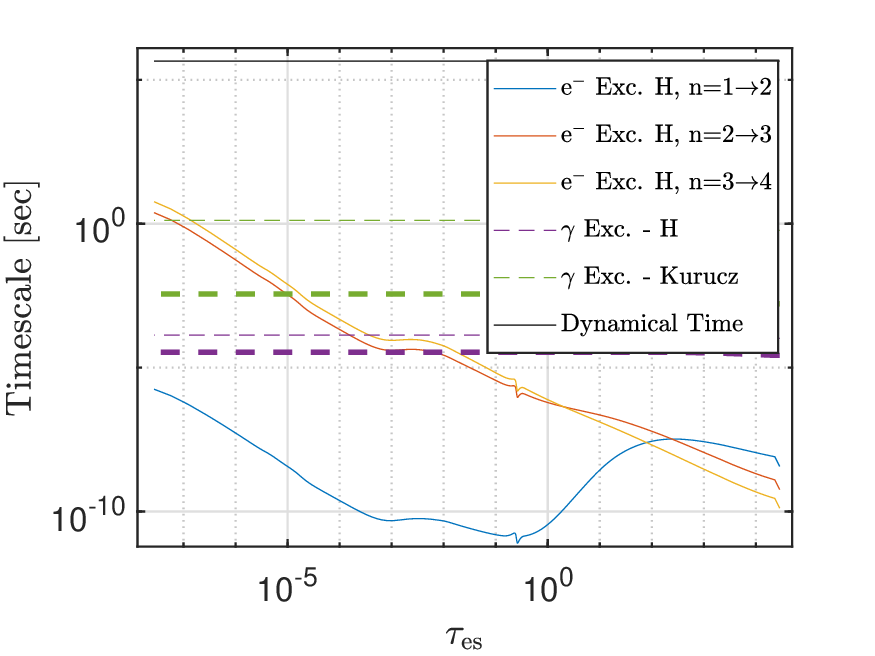}
    \caption{Example relaxation rates in the plasma as function of scattering optical depth $\tau_{\rm es}$ shown at recombination time, $t_{0.7 \rm \, eV}$, when relaxation rates are slowest. For clarity, electron collisional (photon) processes are shown in solid (dashed) lines. In the top panel,  time scales for ionization and for velocity distribution thermalization by Coulomb interactions are shown. At bottom, electron collisional excitation and photoexcitation time scales are shown. Thin (thick) dashed lines denote maximum (average) photon excitation timescales based on an average of the atomic lines (see text). The characteristic time scales of all processes, except electron collisional ionization, are much shorter than the dynamical time.}
    \label{fig:ion_exc_rates}
\end{figure}

The relatively large density implies that the time scale for all relevant processes (electron-electron and electron-ion collisions, photo-ionization and excitation, electron impact excitation), with the exception of electron impact ionization, are much shorter than the dynamical time (see fig.~\ref{fig:ion_exc_rates}). This, combined with the fact that the photon spectrum at energies exceeding the Planck peak (e.g. $h\nu \gtrsim 3$ eV during recombination), is close to thermal out to very low $\tau_{\rm es}$, implies that the ionization fraction and the excitation level distribution of the low energy excited states are both close to LTE. The fact that at low optical depth, $\tau_{\rm es} \lesssim 1$, the radiation energy density falls below that of LTE at low photon energy, implies that the level distribution of the higher energy excited states may deviate from LTE at the outer edge of the ejecta. However, we note that the distribution of excited energy levels are strongly dominated by UV transitions to and from the highly populated ground state \citep[accounting for roughly half the rate, e.g.][]{zeldovich_physics_2002}, hence deviations from LTE occupation of the higher energy states are expected to be mild. In addition, the effect of lines on the SED in this energy range is small, and hence we do not expect a significant effect due to deviations from LTE.

\section{Comparison to earlier works}
\label{sec: Previous works}

\begin{figure}
    \centering
    \includegraphics[width=\columnwidth]{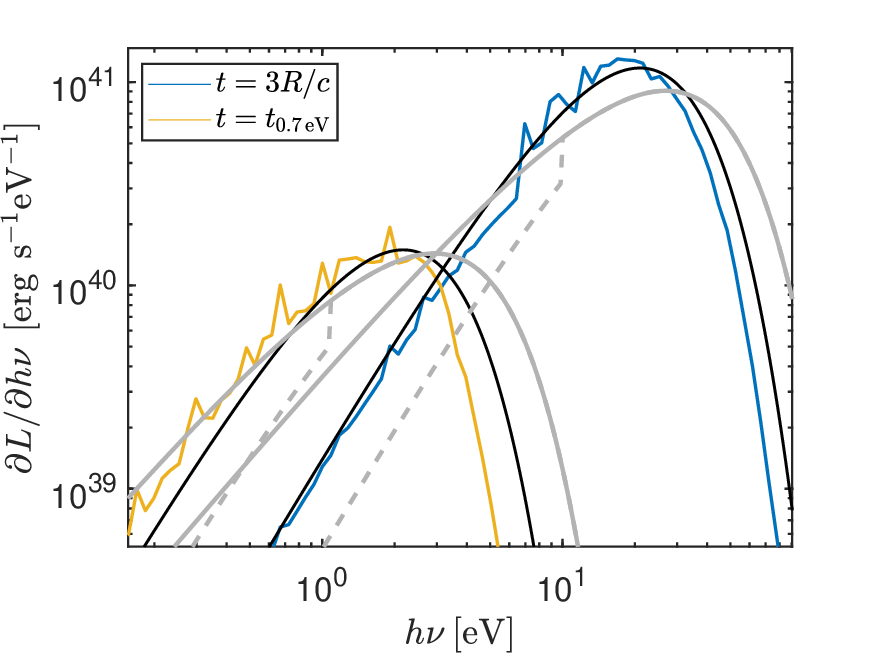}
    \includegraphics[width=\columnwidth]{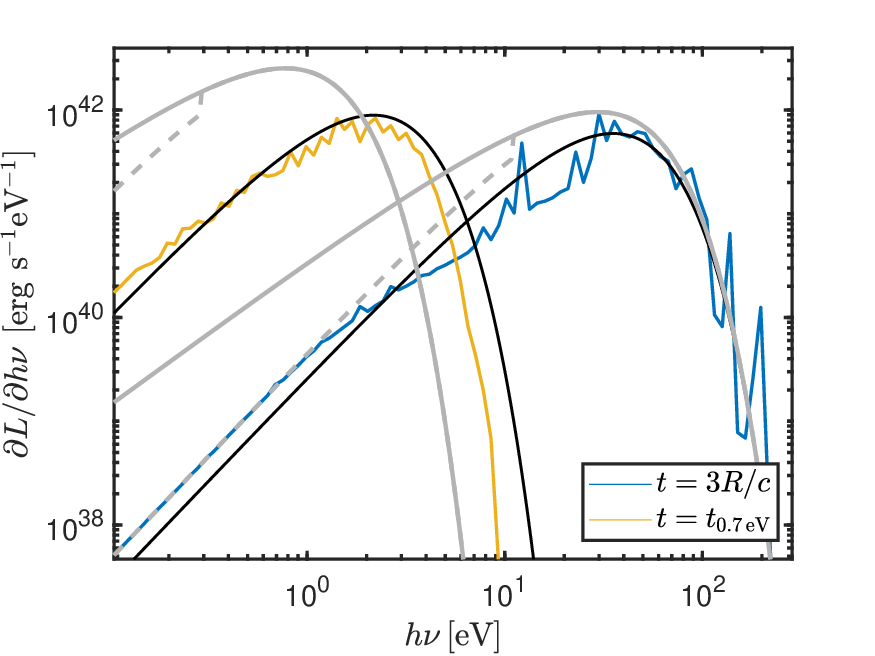}
    \caption{An example of our multigroup simulation results (in color), at two different times, compared with our blackbody formula, eqs. \ref{eq:L_trans}-\ref{eq:L_nu_BB_formula}  (black), and the \citet{shussman_type_2016} formula (solid gray) with the \citet{kozyreva_shock_2020} correction (dashed gray). At top and bottom, explosion energies of $E = 10^{50}$ and $10^{52}$ erg, with progenitor radius $R=10^{13}$ cm, core and envelope masses of $M_{\rm env}=M_{\rm core}=10M_\odot$.}
    \label{fig:us_vs_SWN_vs_Kozyreva}
\end{figure}

\citet{shussman_type_2016} (herafter \citetalias{shussman_type_2016}) produced an analytic model including deviations from blackbody due to free-free opacity. They arrive at an SED that disagrees with our simulations by an order of magnitude or more in the infrared and up to a factor of two near the Planck peak. In \cite{kozyreva_shock_2020}, the \citetalias{shussman_type_2016} formula is compared to simulations produced by the STELLA code. They conclude, similarly, that the \citetalias{shussman_type_2016} model's infrared behavior does not describe the SED for $h\nu<3T_{\rm col}$, and suggest using a blackbody formula for this range. In fig. \ref{fig:us_vs_SWN_vs_Kozyreva}, we compare the \citetalias{shussman_type_2016} formula, including the \citet{kozyreva_shock_2020} corrections, with the results of our MG simulations. We find that their analytic model does not reproduce well the simulation results, due the different values obtained in their analytic approximation for $L$ and $T_{\rm col}$, which in turn is largely due to their opacity approximation neglecting bound bound transitions (see Paper I).

\begin{figure}
    \centering
    \includegraphics[width=\columnwidth]{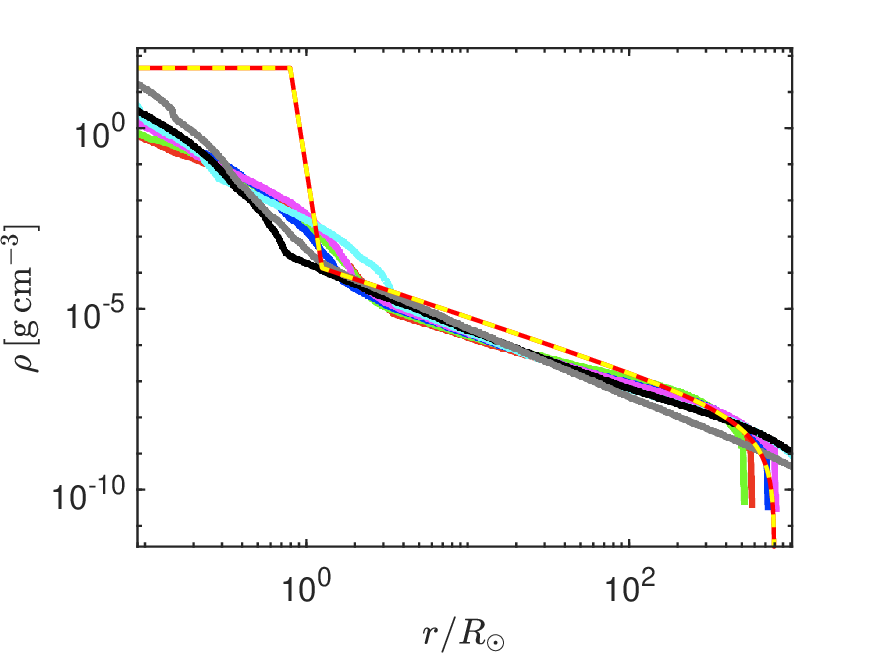}
    \includegraphics[width=\columnwidth]{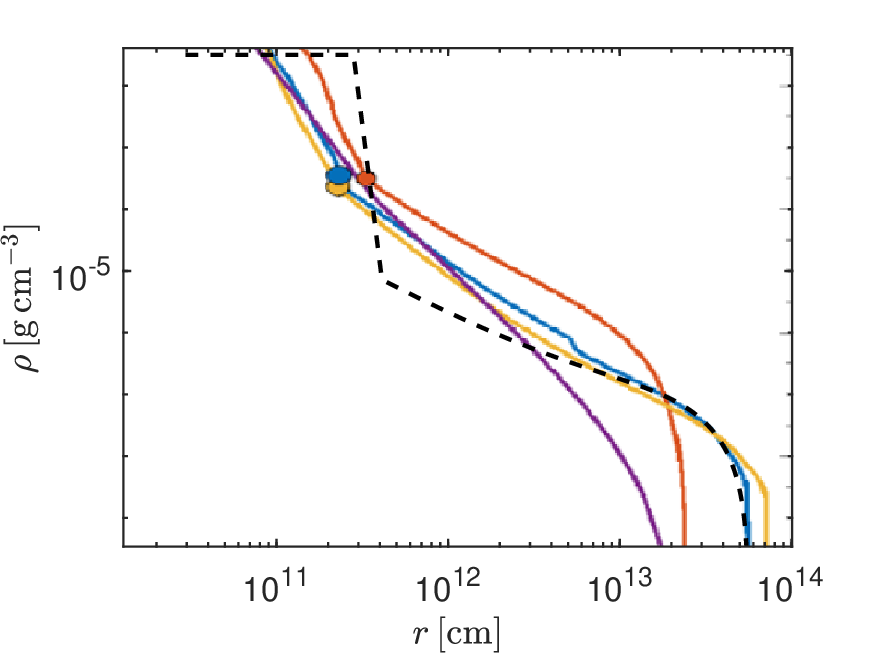}
    \caption{Progenitor density profiles used in earlier shock cooling calculations, compared with the polytropic approximations used in our simulations. \citet{tominaga_shock_2011} at top and \citet{kozyreva_shock_2020} at bottom. The profiles we are approximating are shown in blue, and our approximation profile is shown in dashed lines.}
    \label{fig:Tominaga Kozyerva rho init}
\end{figure}

\begin{figure*}
    \centering
    \includegraphics[width = \textwidth]{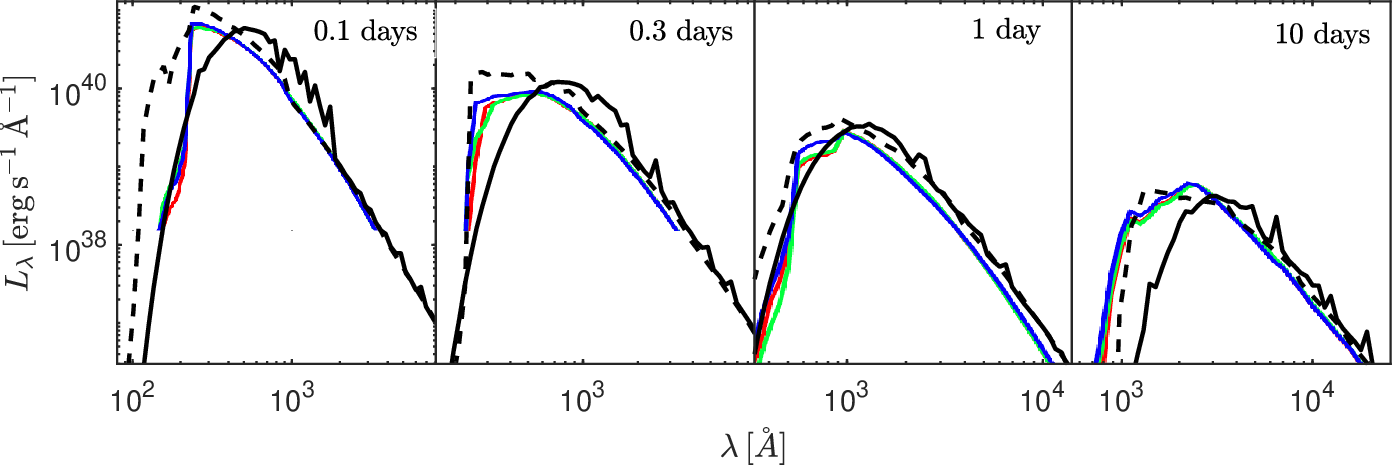}
    \caption{A comparison of the specific shock cooling luminosity, $L_\lambda=\partial L/\partial \lambda$, obtained by \citet{tominaga_shock_2011} in color (different colors represent different choices for the bound-free opacity) and our MG diffusion result in black. Dashed black lines show results for opacity including only bound-free contributions, solid lines show results including both bound-free and bound-bound contributions. See \S~\ref{sec: Previous works} for discussion.}
    \label{fig:Tominaga L_lambda}
\end{figure*}

\begin{figure}
    \centering
    \includegraphics[width=\columnwidth]{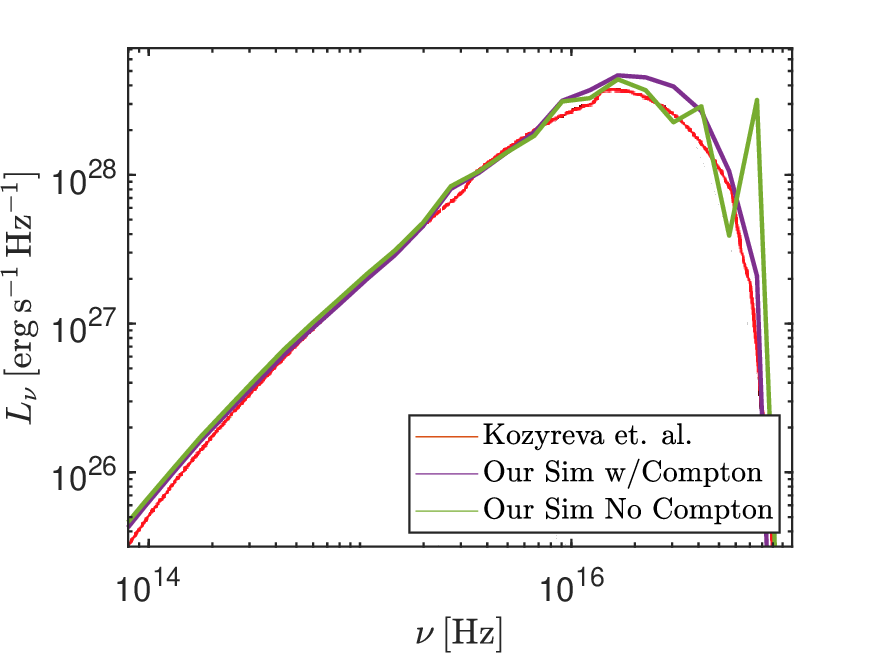}
    \includegraphics[width=\columnwidth]{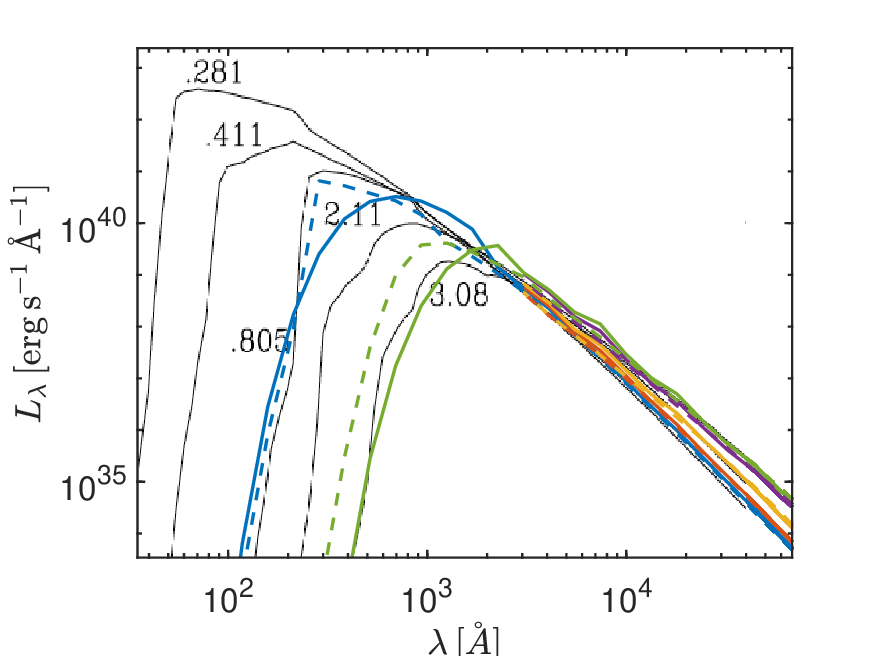}
    \caption{A comparison of the specific luminosity obtained in earlier calculations, \citet{kozyreva_shock_2020} (top) and \citet{blinnikov_comparative_1998} (bottom), with our results. {\it Top}: SED during shock breakout. Our results are shown for calculations with and without inelastic Compton scattering. {\it Bottom}: Shock cooling emission at different times, marked on the curves in units of days. To reduce clutter, we show our results at short wavelengths ($\lambda<3000$\AA) only at $0.281$ days (blue) and $3.08$ days (green). For longer wavelengths, we also include intermediate times (other colors). Dashed (color) lines show results for opacity including only bound-free contributions, solid lines show results including both bound-free and bound-bound contributions. See \S~\ref{sec: Previous works} for discussion.}
    \label{fig:Kozyreva Lbo and Blinnikov}
\end{figure}

Next, we compare our numeric results to those obtained using STELLA \citep{blinnikov_stella_2011} radiation transport calculations \citep{blinnikov_comparative_1998,tominaga_shock_2011,kozyreva_shock_2020} for several non-polytropic profiles obtained using MESA stellar evolution
calculations \citep{paxton_modules_2018}. In each case, we approximate the progenitor density profile used in earlier simulations by a simple polytrope. We show two such examples in fig. \ref{fig:Tominaga Kozyerva rho init} and summarize the profile parameters we use in table \ref{Table:STELLA imitation params}. In figs.~\ref{fig:Tominaga L_lambda} and~\ref{fig:Kozyreva Lbo and Blinnikov} we then compare the resulting emission from our multigroup simulations to the published results, without performing any further fitting. 

The comparison to the results of \citet{tominaga_shock_2011} is shown in fig.~\ref{fig:Tominaga L_lambda}. We find a good agreement with our results when we choose not to include bound-bound opacity. When the line opacity is turned on, as we normally do for our simulations, there is clear disagreement with regards to the peak shape and temperature, though the Rayleigh-Jeans regime shows excellent agreement.

The result in \citet{tominaga_shock_2011} appears to us as non-physical, since the SED peaks at a wavelength corresponding to $\sim3$ times the color temperature, that is expected based on the density and temperature profiles (and on the effective opacity- the scattering optical depth at the location of the corresponding temperature is very high, $\approx100$), and since the SED exhibits a sharp photoionization cutoff, which we do not expect to be observed at this resolution as it should be completely occluded by nearby lines.

The comparison to the results of \cite{kozyreva_shock_2020} is shown in fig. \ref{fig:Kozyreva Lbo and Blinnikov} (top) during shock breakout. We find an excellent agreement between the results of the different calculations (note that at these temperatures only few atomic transition lines contribute).

Finally, fig. \ref{fig:Kozyreva Lbo and Blinnikov} (bottom) shows a comparison to the results of \citet{blinnikov_comparative_1998}. There is reasonable agreement between the results at the latest observed times ($t=3$ days after breakout). At early time, there is good agreement at longer wavelengths, $\lambda>3000$\AA, where the free-free emission dominates, but there is a clear discrepancy at higher frequencies. The very high temperature of the emission peak obtained in \citet{blinnikov_comparative_1998} at these times, $T\sim70$ eV, appears to be non-physical, since similarly to the \citet{tominaga_shock_2011} case, for the corresponding progenitor parameters it is associated with the temperature of the envelope at an optical depth $\tau \sim 1000$. \citet{blinnikov_comparative_1998} arrive at the same conclusion, noting that that particular model features a very weak absorption opacity, and as a result, unreasonably high emission temperatures. 

The agreement of our results with these earlier calculations provides additional support to our code’s validity (in particular to the applicability of the diffusion approximation), and to the conclusion of the detailed analysis of SW17, that the shock cooling emission is not sensitive to deviations of the density profile from a polytropic one.

\begin{table*}
\centering
\begin{tabular}{ |p{2.6cm}||p{2.6cm}|p{2.3cm}|p{2.4cm}|p{2.6cm}| }
\hline
Previous Work &  Progenitor Radius [cm] & Core Mass [$M_\odot$] & Envelope Mass [$M_\odot$] & Explosion Energy [erg] \\
\hline
 \citet{blinnikov_comparative_1998}   & $4.33\times10^{13}$  & 2.2 & 0.2  & $1.2\times10^{51}$ \\
 \citet{tominaga_shock_2011} &  $5.5\times10^{13}$   & 15 & 2 & $10^{51}$ \\
 \citet{kozyreva_shock_2020} & $5.6\times10^{13}$ & 1.5 & 7.5 & $10^{51}$ \\
 \hline
\end{tabular}
\caption{Parameters of the polytropic models that we use to approximate existing STELLA simulations in the literature. 
}
\label{Table:STELLA imitation params}
\end{table*}

\section{Discussion and Summary}
\label{sec: Summary}

We derived an analytic description of the deviations of shock cooling spectra from blackbody for red supergiant explosions. The approximation is given by eqs.~(\ref{eq:Lnu epsilon final}) and (\ref{eq:Lnu epsilon simplified}). The definitions of (and equations for) all variables appearing in eqs.~(\ref{eq:Lnu epsilon final}) and~(\ref{eq:Lnu epsilon simplified}) are given in the Appendix. The analytic description holds from post breakout ($t>3R/c$), up to H recombination ($T\approx0.7$~eV), or until the photon diffusion time through the envelope becomes shorter than the dynamical time, see eq.~(\ref{eq:t_transp}). The analytic expressions were calibrated against a large set of numeric MG calculations, and provide an excellent approximation to the numeric results- see \S~\ref{sec:numeric_res}, figures ~\ref{fig:Lnu_tiles_lowR} and~\ref{fig:Lnu_tiles_hiR}. In accordance with \citetalias{sapir_uv/optical_2017} and \citetalias{morag_shock_2023}, we find that the results are not sensitive to metalicity and to the ratio of core to stellar radii, in the range $R_{\rm c}/R=10^{-1}-10^{-3}$. In \S~\ref{sec: finite frequency resolution} and \S~\ref{sec:NLTE} we showed the effects of deviations from ionization and excitation LTE and of `expansion opacity' corrections are small.

Our analytic formula depends on the same four parameters as the previous one in \citetalias{morag_shock_2023}, $\{R,{\rm v_{\rm s\ast}},f_\rho M,M_{\rm env}\}$. Of these parameters, the SED is most sensitive to the progenitor radius $R$, and the ejecta velocity $\rm v_{\rm s\ast}$, and insensitive to the parameters $f_\rho M$ and $M_{\rm env}$, where $M$ and $M_{\rm env}$ are the total mass of the ejecta and mass of the envelope, and $f_{\rho}$ is a dimensionless factor of order unity that depends on the inner envelope structure (see eq.~\ref{eq:rho_in}). Therefore, the former two parameters are the ones most readily extracted from observations. We note also that deducing parameter values is hindered by the difficulty of determining $T_{\rm col}$ at early times, when the maximum observable photon frequencies ($h\nu\lesssim$~10 eV) may be located below the thermal peak. For this regime, the frequency-dependent deviations from blackbody may prove to be an important discriminator between models of the Raleigh-Jeans part of the spectrum.

Our results are compared in \S~\ref{sec: Previous works} to earlier numerical results including STELLA radiation transport calculation results for several non-polytropic profiles \citep{blinnikov_comparative_1998,tominaga_shock_2011,kozyreva_shock_2020} obtained using MESA  stellar evolution calculations. The agreement of our results with these earlier calculations provides additional support to our code's validity (in particular to the applicability of the diffusion approximation), and to the conclusion of the detailed analysis of SW17, that the shock cooling emission is not sensitive to deviations of the density profile from a polytropic one. 

Based on our analysis of \S~\ref{sec: finite frequency resolution}, we find that the emergence of Balmer lines does not coincide with the recombination time but likely occurs earlier (approximately at $T\sim2-3$ eV). While the recombined H fraction at this time is very low, and the effect of the lines on the SED should be small, the lines may be visible in a spectrum. The appearance of the first Balmer lines should not be associated with the recombination time $t_{0.7 \rm \, eV}$, when the temperature drops to 0.7~eV.

Our frequency-dependent formula agrees with the simulations' results with an RMS error of $\lesssim20\%$ (with the exception of the Wien tail, where deviations are larger but reflect a very small inaccuracy in radiation temperature). This error is somewhat larger than the uncertainty in our numeric results, $\approx 10\%$, as quantified by the difference between the simulations' results and the high frequency-resolution calculation (eq. \ref{eq: Doppler Integral}). Therefore, in order to best incorporate the uncertainty of our analytic model when comparing to observations, we suggest using a covariance matrix (as function of $(\nu,t)$) of the residuals between the frequency dependent formula (eq. \ref{eq:Lnu epsilon final} or \ref{eq:Lnu epsilon simplified}) and our published simulations (see \S Data Availability).

\section*{Acknowledgements}
We thank Barack Zackay for his contribution  to our numerical code, as well as Gilad Sadeh for insightful discussion. EW's research is partially supported by ISF, GIF and IMOS grants.

\section*{Data Availability}
Numerical codes used in this paper will be provided upon reasonable request to the corresponding author. Our opacity table code is available online for public use at \cite{morag_frequency_2023}. Numeric results from our simulations are available at \url{https://www.dropbox.com/s/ub5zg1rngodjb1d/RSG_Solar_Metallicity.mat?dl=0}.

\clearpage

\bibliographystyle{mnras}
\bibliography{references}

\begin{thebibliography}{}
\makeatletter
\relax
\def\mn@urlcharsother{\let\do\@makeother \do\$\do\&\do\#\do\^\do\_\do\%\do\~}
\def\mn@doi{\begingroup\mn@urlcharsother \@ifnextchar [ {\mn@doi@} {\mn@doi@[]}}
\def\mn@doi@[#1]#2{\def\@tempa{#1}\ifx\@tempa\@empty \href {http://dx.doi.org/#2} {doi:#2}\else \href {http://dx.doi.org/#2} {#1}\fi \endgroup}
\def\mn@eprint#1#2{\mn@eprint@#1:#2::\@nil}
\def\mn@eprint@arXiv#1{\href {http://arxiv.org/abs/#1} {{\tt arXiv:#1}}}
\def\mn@eprint@dblp#1{\href {http://dblp.uni-trier.de/rec/bibtex/#1.xml} {dblp:#1}}
\def\mn@eprint@#1:#2:#3:#4\@nil{\def\@tempa {#1}\def\@tempb {#2}\def\@tempc {#3}\ifx \@tempc \@empty \let \@tempc \@tempb \let \@tempb \@tempa \fi \ifx \@tempb \@empty \def\@tempb {arXiv}\fi \@ifundefined {mn@eprint@\@tempb}{\@tempb:\@tempc}{\expandafter \expandafter \csname mn@eprint@\@tempb\endcsname \expandafter{\@tempc}}}

\bibitem[\protect\citeauthoryear{Baron, Hauschildt, Nugent  \& Branch}{Baron et~al.}{1996}]{baron_non-local_1996}
Baron E.,  Hauschildt P.~H.,  Nugent P.,   Branch D.,  1996, \mn@doi [Monthly Notices of the Royal Astronomical Society] {10.1093/mnras/283.1.297}, 283, 297

\bibitem[\protect\citeauthoryear{Baron et~al.,}{Baron et~al.}{2000}]{baron_preliminary_2000}
Baron E.,  et~al., 2000, \mn@doi [The Astrophysical Journal] {10.1086/317795}, 545, 444

\bibitem[\protect\citeauthoryear{Baron, Nugent, Branch, Hauschildt, Turatto  \& Cappellaro}{Baron et~al.}{2003}]{baron_determination_2003}
Baron E.,  Nugent P.~E.,  Branch D.,  Hauschildt P.~H.,  Turatto M.,   Cappellaro E.,  2003, \mn@doi [The Astrophysical Journal] {10.1086/367888}, 586, 1199

\bibitem[\protect\citeauthoryear{Blinnikov \& Bartunov}{Blinnikov \& Bartunov}{2011}]{blinnikov_stella_2011}
Blinnikov S.~I.,  Bartunov O.~S.,  2011, Astrophysics Source Code Library, p. ascl:1108.013

\bibitem[\protect\citeauthoryear{Blinnikov, Eastman, Bartunov, Popolitov  \& Woosley}{Blinnikov et~al.}{1998}]{blinnikov_comparative_1998}
Blinnikov S.~I.,  Eastman R.,  Bartunov O.~S.,  Popolitov V.~A.,   Woosley S.~E.,  1998, \mn@doi [The Astrophysical Journal] {10.1086/305375}, 496, 454

\bibitem[\protect\citeauthoryear{Blinnikov, Lundqvist, Bartunov, Nomoto  \& Iwamoto}{Blinnikov et~al.}{2000}]{blinnikov_radiation_2000}
Blinnikov S.,  Lundqvist P.,  Bartunov O.,  Nomoto K.,   Iwamoto K.,  2000, \mn@doi [The Astrophysical Journal] {10.1086/308588}, 532, 1132

\bibitem[\protect\citeauthoryear{Calzavara \& Matzner}{Calzavara \& Matzner}{2004}]{calzavara_supernova_2004}
Calzavara A.~J.,  Matzner C.~D.,  2004, \mn@doi [Monthly Notices of the Royal Astronomical Society] {10.1111/j.1365-2966.2004.07818.x}, 351, 694

\bibitem[\protect\citeauthoryear{Castor}{Castor}{2007}]{castor_radiation_2007}
Castor J.~I.,  2007, Radiation {Hydrodynamics}.
Cambridge, UK ; New York

\bibitem[\protect\citeauthoryear{Chevalier}{Chevalier}{1992}]{chevalier_early_1992}
Chevalier R.~A.,  1992, \mn@doi [The Astrophysical Journal] {10.1086/171612}, 394, 599

\bibitem[\protect\citeauthoryear{Colgan, Kilcrease, Magee, Abdallah, Sherrill, Fontes, Hakel  \& Zhang}{Colgan et~al.}{2015}]{colgan_light_2015}
Colgan J.,  Kilcrease D.~P.,  Magee N.~H.,  Abdallah J.,  Sherrill M.~E.,  Fontes C.~J.,  Hakel P.,   Zhang H.~L.,  2015, \mn@doi [High Energy Density Physics] {10.1016/j.hedp.2015.02.006}, 14, 33

\bibitem[\protect\citeauthoryear{Colgan et~al.,}{Colgan et~al.}{2016}]{colgan_new_2016}
Colgan J.,  et~al., 2016, \mn@doi [The Astrophysical Journal] {10.3847/0004-637X/817/2/116}, 817, 116

\bibitem[\protect\citeauthoryear{Colgan, Kilcrease, Magee, Sherrill, Fontes  \& Hakel}{Colgan et~al.}{2018}]{colgan_new_2018}
Colgan J.,  Kilcrease D.~P.,  Magee N.~H.,  Sherrill M.~E.,  Fontes C.~J.,   Hakel P.,  2018, \mn@doi [Atoms] {10.3390/atoms6020032}, 6, 32

\bibitem[\protect\citeauthoryear{Dappen, Anderson  \& Mihalas}{Dappen et~al.}{1987}]{dappen_statistical_1987}
Dappen W.,  Anderson L.,   Mihalas D.,  1987, \mn@doi [The Astrophysical Journal] {10.1086/165446}, 319, 195

\bibitem[\protect\citeauthoryear{Dessart \& Hillier}{Dessart \& Hillier}{2005}]{dessart_distance_2005}
Dessart L.,  Hillier D.~J.,  2005, \mn@doi [Astronomy \& Astrophysics] {10.1051/0004-6361:20053217}, 439, 671

\bibitem[\protect\citeauthoryear{Dessart \& Hillier}{Dessart \& Hillier}{2008}]{dessart_time-dependent_2008}
Dessart L.,  Hillier D.~J.,  2008, \mn@doi [Monthly Notices of the Royal Astronomical Society] {10.1111/j.1365-2966.2007.12538.x}, 383, 57

\bibitem[\protect\citeauthoryear{Dessart, Audit  \& Hillier}{Dessart et~al.}{2015}]{dessart_numerical_2015}
Dessart L.,  Audit E.,   Hillier D.~J.,  2015, \mn@doi [Monthly Notices of the Royal Astronomical Society] {10.1093/mnras/stv609}, 449, 4304

\bibitem[\protect\citeauthoryear{Eastman \& Pinto}{Eastman \& Pinto}{1993}]{Eastman_Spectrum_1993}
Eastman R.~G.,  Pinto P.~A.,  1993, \mn@doi [The Astrophysical Journal] {10.1086/172957}, 412, 731

\bibitem[\protect\citeauthoryear{Ergon, Fransson, Jerkstrand, Kozma, Kromer  \& Spricer}{Ergon et~al.}{2018}]{ergon_monte-carlo_2018}
Ergon M.,  Fransson C.,  Jerkstrand A.,  Kozma C.,  Kromer M.,   Spricer K.,  2018, \mn@doi [Astronomy \& Astrophysics] {10.1051/0004-6361/201833043}, 620, A156

\bibitem[\protect\citeauthoryear{Friend \& Castor}{Friend \& Castor}{1983}]{friend_stellar_1983}
Friend D.~B.,  Castor J.~I.,  1983, \mn@doi [The Astrophysical Journal] {10.1086/161289}, 272, 259

\bibitem[\protect\citeauthoryear{Förster et~al.,}{Förster et~al.}{2018}]{forster_delay_2018}
Förster F.,  et~al., 2018, \mn@doi [Nature Astronomy] {10.1038/s41550-018-0563-4}, 2, 808

\bibitem[\protect\citeauthoryear{Gal-Yam et~al.,}{Gal-Yam et~al.}{2011}]{gal-yam_real-time_2011}
Gal-Yam A.,  et~al., 2011, \mn@doi [The Astrophysical Journal] {10.1088/0004-637X/736/2/159}, 736, 159

\bibitem[\protect\citeauthoryear{Gandel'Man \& Frank-Kamenetskii}{Gandel'Man \& Frank-Kamenetskii}{1956}]{gandelman_shock_1956}
Gandel'Man G.~M.,  Frank-Kamenetskii D.~A.,  1956, Soviet Physics Doklady, 1, 223

\bibitem[\protect\citeauthoryear{Gasilov, Kuchugov, Olkhovskaya  \& Chetverushkin}{Gasilov et~al.}{2016}]{gasilov_solution_2016}
Gasilov V.~A.,  Kuchugov P.~A.,  Olkhovskaya O.~G.,   Chetverushkin B.~N.,  2016, \mn@doi [Computational Mathematics and Mathematical Physics] {10.1134/S0965542516060130}, 56, 987

\bibitem[\protect\citeauthoryear{Hummer \& Mihalas}{Hummer \& Mihalas}{1988}]{hummer_equation_1988}
Hummer D.~G.,  Mihalas D.,  1988, \mn@doi [The Astrophysical Journal] {10.1086/166600}, 331, 794

\bibitem[\protect\citeauthoryear{Iglesias \& Rogers}{Iglesias \& Rogers}{1996}]{iglesias_updated_1996}
Iglesias C.,  Rogers F.,  1996, Astrophysical Journal, 464, 943

\bibitem[\protect\citeauthoryear{Irani et~al.,}{Irani et~al.}{2022}]{irani_sn_2022}
Irani I.,  et~al., 2022, {SN} 2022oqm -- a {Ca}-rich explosion of a compact progenitor embedded in {C}/{O} circumstellar material, \mn@doi{10.48550/arXiv.2210.02554}, \url {http://arxiv.org/abs/2210.02554}

\bibitem[\protect\citeauthoryear{Ivezić et~al.,}{Ivezić et~al.}{2019}]{ivezic_lsst_2019}
Ivezić Z.,  et~al., 2019, \mn@doi [The Astrophysical Journal] {10.3847/1538-4357/ab042c}, 873, 111

\bibitem[\protect\citeauthoryear{Jacobson-Galán et~al.,}{Jacobson-Galán et~al.}{2022}]{jacobson-galan_circumstellar_2022}
Jacobson-Galán W.,  et~al., 2022, The {Circumstellar} {Environments} of {Double}-{Peaked}, {Calcium}-strong {Supernovae} 2021gno and 2021inl, \url {https://ui.adsabs.harvard.edu/abs/2022arXiv220303785J}

\bibitem[\protect\citeauthoryear{Katz, Sapir  \& Waxman}{Katz et~al.}{2012}]{katz_non-relativistic_2012}
Katz B.,  Sapir N.,   Waxman E.,  2012, \mn@doi [The Astrophysical Journal] {10.1088/0004-637X/747/2/147}, 747, 147

\bibitem[\protect\citeauthoryear{Kerzendorf \& Sim}{Kerzendorf \& Sim}{2014}]{kerzendorf_spectral_2014}
Kerzendorf W.~E.,  Sim S.~A.,  2014, \mn@doi [Monthly Notices of the Royal Astronomical Society] {10.1093/mnras/stu055}, 440, 387

\bibitem[\protect\citeauthoryear{Kleiser \& Kasen}{Kleiser \& Kasen}{2012}]{kleiser_fast_2012}
Kleiser I.,  Kasen D.,  2012. p. 242.29, \url {http://adsabs.harvard.edu/abs/2012AAS...21924229K}

\bibitem[\protect\citeauthoryear{Kozyreva, Nakar, Waldman, Blinnikov  \& Baklanov}{Kozyreva et~al.}{2020}]{kozyreva_shock_2020}
Kozyreva A.,  Nakar E.,  Waldman R.,  Blinnikov S.,   Baklanov P.,  2020, \mn@doi [Monthly Notices of the Royal Astronomical Society] {10.1093/mnras/staa924}, 494, 3927

\bibitem[\protect\citeauthoryear{Kromer \& Sim}{Kromer \& Sim}{2009}]{kromer_time-dependent_2009}
Kromer M.,  Sim S.~A.,  2009, \mn@doi [Monthly Notices of the Royal Astronomical Society] {10.1111/j.1365-2966.2009.15256.x}, 398, 1809

\bibitem[\protect\citeauthoryear{Kurucz}{Kurucz}{1995}]{kurucz_atomic_1995}
Kurucz R.~L.,  1995, in {ASP} {Conference} {Series}. p.~583, \url {http://adsabs.harvard.edu/abs/1995ASPC...81..583K}

\bibitem[\protect\citeauthoryear{Levinson \& Nakar}{Levinson \& Nakar}{2020}]{levinson_physics_2020}
Levinson A.,  Nakar E.,  2020, \mn@doi [Physics Reports] {10.1016/j.physrep.2020.04.003}, 866, 1

\bibitem[\protect\citeauthoryear{Lisakov, Dessart, Hillier, Waldman  \& Livne}{Lisakov et~al.}{2017}]{lisakov_study_2017}
Lisakov S.~M.,  Dessart L.,  Hillier D.~J.,  Waldman R.,   Livne E.,  2017, \mn@doi [Monthly Notices of the Royal Astronomical Society] {10.1093/mnras/stw3035}, 466, 34

\bibitem[\protect\citeauthoryear{Matzner \& McKee}{Matzner \& McKee}{1999}]{matzner_expulsion_1999}
Matzner C.~D.,  McKee C.~F.,  1999, \mn@doi [The Astrophysical Journal] {10.1086/306571}, 510, 379

\bibitem[\protect\citeauthoryear{McClarren, Evans, Lowrie  \& Densmore}{McClarren et~al.}{2008}]{mcclarren_semi-implicit_2008}
McClarren R.~G.,  Evans T.~M.,  Lowrie R.~B.,   Densmore J.~D.,  2008, \mn@doi [Journal of Computational Physics] {10.1016/j.jcp.2008.04.029}, 227, 7561

\bibitem[\protect\citeauthoryear{Mihalas \& Mihalas}{Mihalas \& Mihalas}{1999}]{mihalas_foundations_1999}
Mihalas D.,  Mihalas B.~W.,  1999, Foundations of {Radiation} {Hydrodynamics}.
Mineola, NY

\bibitem[\protect\citeauthoryear{Morag}{Morag}{2023}]{morag_frequency_2023}
Morag J.,  2023, Frequency {Dependent} {Opacity} {Table}, \url {https://github.com/jon-morag/Freq_Dept_Opac_Table/}

\bibitem[\protect\citeauthoryear{Morag, Sapir  \& Waxman}{Morag et~al.}{2023}]{morag_shock_2023}
Morag J.,  Sapir N.,   Waxman E.,  2023, \mn@doi [Monthly Notices of the Royal Astronomical Society] {10.1093/mnras/stad899}, 522, 2764

\bibitem[\protect\citeauthoryear{Nakar \& Sari}{Nakar \& Sari}{2010}]{nakar_early_2010}
Nakar E.,  Sari R.,  2010, \mn@doi [The Astrophysical Journal] {10.1088/0004-637X/725/1/904}, 725, 904

\bibitem[\protect\citeauthoryear{Ohyama}{Ohyama}{1963}]{ohyama_explosion_1963}
Ohyama N.,  1963, \mn@doi [Progress of Theoretical Physics] {10.1143/PTP.30.170}, 30, 170

\bibitem[\protect\citeauthoryear{Paxton et~al.,}{Paxton et~al.}{2018}]{paxton_modules_2018}
Paxton B.,  et~al., 2018, \mn@doi [The Astrophysical Journal Supplement Series] {10.3847/1538-4365/aaa5a8}, 234, 34

\bibitem[\protect\citeauthoryear{Pinto \& Eastman}{Pinto \& Eastman}{2000}]{pinto_physics_2000}
Pinto P.~A.,  Eastman R.~G.,  2000, \mn@doi [The Astrophysical Journal] {10.1086/308380}, 530, 757

\bibitem[\protect\citeauthoryear{Piro, Muhleisen, Arcavi, Sand, Tartaglia  \& Valenti}{Piro et~al.}{2017}]{piro_numerically_2017}
Piro A.~L.,  Muhleisen M.,  Arcavi I.,  Sand D.~J.,  Tartaglia L.,   Valenti S.,  2017, \mn@doi [The Astrophysical Journal] {10.3847/1538-4357/aa8595}, 846, 94

\bibitem[\protect\citeauthoryear{Rabinak \& Waxman}{Rabinak \& Waxman}{2011}]{rabinak_early_2011}
Rabinak I.,  Waxman E.,  2011, \mn@doi [The Astrophysical Journal] {10.1088/0004-637X/728/1/63}, 728, 63

\bibitem[\protect\citeauthoryear{Roth \& Kasen}{Roth \& Kasen}{2015}]{roth_monte_2015}
Roth N.,  Kasen D.,  2015, \mn@doi [The Astrophysical Journal Supplement Series] {10.1088/0067-0049/217/1/9}, 217, 9

\bibitem[\protect\citeauthoryear{Rubin \& Gal-Yam}{Rubin \& Gal-Yam}{2017}]{rubin_exploring_2017}
Rubin A.,  Gal-Yam A.,  2017, \mn@doi [The Astrophysical Journal] {10.3847/1538-4357/aa8465}, 848, 8

\bibitem[\protect\citeauthoryear{Rybicki \& Lightman}{Rybicki \& Lightman}{1979}]{rybicki_radiative_1979}
Rybicki G.~B.,  Lightman A.~P.,  1979, Radiative processes in astrophysics.
\url {https://ui.adsabs.harvard.edu/abs/1979rpa..book.....R}

\bibitem[\protect\citeauthoryear{Sagiv et~al.,}{Sagiv et~al.}{2014}]{sagiv_science_2014}
Sagiv I.,  et~al., 2014, \mn@doi [The Astronomical Journal] {10.1088/0004-6256/147/4/79}, 147, 79

\bibitem[\protect\citeauthoryear{Sakurai}{Sakurai}{1960}]{sakurai_problem_1960}
Sakurai A.,  1960, COMMUNICATIONS ON PURE AND APPLIED MATHEMATIC, XIII, 353

\bibitem[\protect\citeauthoryear{Sapir \& Halbertal}{Sapir \& Halbertal}{2014}]{sapir_numeric_2014}
Sapir N.,  Halbertal D.,  2014, \mn@doi [The Astrophysical Journal] {10.1088/0004-637X/796/2/145}, 796, 145

\bibitem[\protect\citeauthoryear{Sapir \& Waxman}{Sapir \& Waxman}{2017}]{sapir_uv/optical_2017}
Sapir N.,  Waxman E.,  2017, \mn@doi [The Astrophysical Journal] {10.3847/1538-4357/aa64df}, 838, 130

\bibitem[\protect\citeauthoryear{Sapir, Katz  \& Waxman}{Sapir et~al.}{2011}]{sapir_non-relativistic_2011}
Sapir N.,  Katz B.,   Waxman E.,  2011, \mn@doi [The Astrophysical Journal] {10.1088/0004-637X/742/1/36}, 742, 36

\bibitem[\protect\citeauthoryear{Sapir, Katz  \& Waxman}{Sapir et~al.}{2013}]{sapir_non-relativistic_2013}
Sapir N.,  Katz B.,   Waxman E.,  2013, \mn@doi [The Astrophysical Journal] {10.1088/0004-637X/774/1/79}, 774, 79

\bibitem[\protect\citeauthoryear{Shussman, Waldman  \& Nakar}{Shussman et~al.}{2016}]{shussman_type_2016}
Shussman T.,  Waldman R.,   Nakar E.,  2016, arXiv e-prints, p. arXiv:1610.05323

\bibitem[\protect\citeauthoryear{Shvartzvald et~al.,}{Shvartzvald et~al.}{2023}]{shvartzvald_ultrasat_2023}
Shvartzvald Y.,  et~al., 2023, {ULTRASAT}: {A} wide-field time-domain {UV} space telescope, \mn@doi{10.48550/arXiv.2304.14482}, \url {http://arxiv.org/abs/2304.14482}

\bibitem[\protect\citeauthoryear{Skartlien}{Skartlien}{2000}]{skartlien_multigroup_2000}
Skartlien R.,  2000, \mn@doi [Astrophysical Journal] {10.1086/308934}, 536, 465

\bibitem[\protect\citeauthoryear{Takeda}{Takeda}{1990}]{takeda_non-lte_1990}
Takeda Y.,  1990, Astronomy and Astrophysics, 234, 343

\bibitem[\protect\citeauthoryear{Takeda}{Takeda}{1991}]{takeda_formation_1991}
Takeda Y.,  1991, Astronomy and Astrophysics, 245, 182

\bibitem[\protect\citeauthoryear{Tominaga, Blinnikov, Baklanov, Morokuma, Nomoto  \& Suzuki}{Tominaga et~al.}{2009}]{tominaga_properties_2009}
Tominaga N.,  Blinnikov S.,  Baklanov P.,  Morokuma T.,  Nomoto K.,   Suzuki T.,  2009, \mn@doi [The Astrophysical Journal] {10.1088/0004-637X/705/1/L10}, 705, L10

\bibitem[\protect\citeauthoryear{Tominaga, Morokuma, Blinnikov, Baklanov, Sorokina  \& Nomoto}{Tominaga et~al.}{2011}]{tominaga_shock_2011}
Tominaga N.,  Morokuma T.,  Blinnikov S.~I.,  Baklanov P.,  Sorokina E.~I.,   Nomoto K.,  2011, \mn@doi [The Astrophysical Journal Supplement Series] {10.1088/0067-0049/193/1/20}, 193, 20

\bibitem[\protect\citeauthoryear{Tominaga, Shibata  \& Blinnikov}{Tominaga et~al.}{2015}]{tominaga_time-dependent_2015}
Tominaga N.,  Shibata S.,   Blinnikov S.~I.,  2015, \mn@doi [The Astrophysical Journal Supplement Series] {10.1088/0067-0049/219/2/38}, 219, 38

\bibitem[\protect\citeauthoryear{Tsang, Goldberg, Bildsten  \& Kasen}{Tsang et~al.}{2020}]{tsang_comparing_2020}
Tsang B. T.-H.,  Goldberg J.~A.,  Bildsten L.,   Kasen D.,  2020, \mn@doi [The Astrophysical Journal] {10.3847/1538-4357/ab989d}, 898, 29

\bibitem[\protect\citeauthoryear{Utrobin}{Utrobin}{2007}]{utrobin_optimal_2007}
Utrobin V.~P.,  2007, \mn@doi [Astronomy and Astrophysics] {10.1051/0004-6361:20066078}, 461, 233

\bibitem[\protect\citeauthoryear{Utrobin \& Chugai}{Utrobin \& Chugai}{2005}]{utrobin_strong_2005}
Utrobin V.~P.,  Chugai N.~N.,  2005, \mn@doi [Astronomy \& Astrophysics] {10.1051/0004-6361:20042599}, 441, 271

\bibitem[\protect\citeauthoryear{Vaytet, Audit, Dubroca  \& Delahaye}{Vaytet et~al.}{2011}]{vaytet_numerical_2011}
Vaytet N. M.~H.,  Audit E.,  Dubroca B.,   Delahaye F.,  2011, \mn@doi [Journal of Quantitative Spectroscopy and Radiative Transfer] {10.1016/j.jqsrt.2011.01.027}, 112, 1323

\bibitem[\protect\citeauthoryear{Verner, Ferland, Korista  \& Yakovlev}{Verner et~al.}{1996}]{verner_atomic_1996}
Verner D.~A.,  Ferland G.~J.,  Korista K.~T.,   Yakovlev D.~G.,  1996, \mn@doi [The Astrophysical Journal] {10.1086/177435}, 465, 487

\bibitem[\protect\citeauthoryear{Waxman \& Katz}{Waxman \& Katz}{2016}]{waxman_shock_2016}
Waxman E.,  Katz B.,  2016, arXiv:1607.01293 [astro-ph]

\bibitem[\protect\citeauthoryear{Waxman, Ofek, Kushnir  \& Gal-Yam}{Waxman et~al.}{2018}]{waxman_constraints_2018}
Waxman E.,  Ofek E.~O.,  Kushnir D.,   Gal-Yam A.,  2018, \mn@doi [Monthly Notices of the Royal Astronomical Society] {10.1093/mnras/sty2441}, 481, 3423

\bibitem[\protect\citeauthoryear{Weaver}{Weaver}{1976}]{weaver_structure_1976}
Weaver T.~A.,  1976, \mn@doi [The Astrophysical Journal Supplement Series] {10.1086/190398}, 32, 233

\bibitem[\protect\citeauthoryear{Zel’dovich \& Raizer}{Zel’dovich \& Raizer}{2002}]{zeldovich_physics_2002}
Zel’dovich Y.~B.,  Raizer Y.~P.,  2002, Physics of {Shock} {Waves} and {High}-{Temperature} {Hydrodynamic} {Phenomena}, annotated edition edn.
Dover Publications, Mineola, N.Y

\makeatother
\end{thebibliography}

\appendix

\section{Summary of model equations}
\label{appendix}

The basis for the analytic approximations of the spectra given in this paper  is the analytic approximation of Paper I for the bolomteric luminosity 
$L$  and the color temperature $T_{\rm col}$ (eqs. \ref{eq:L_trans}-\ref{eq:T_trans}),
\begin{equation}
    L/L_{\rm br}=\tilde{t}^{-4/3}+\tilde{t}^{-0.172}\times A\exp\left(-\left[at/t_{\rm tr}\right]^{\alpha}\right), \\
    \label{eq:L_trans_Appendix}
\end{equation}
\begin{equation}
    T_{\rm col}/T_{\rm col,br}=\min\left[0.97\,\tilde{t}^{-1/3},\tilde{t}^{-0.45}\right].
    \label{eq:T_trans_Appendix}
\end{equation}
Here $\{A,a,\alpha\} =\{0.9,2,0.5\}$, $\tilde{t}=t/t_{\rm br}$, and we define $t=0$ as the time at which the breakout flux peaks. The break parameters (with br subscript) are given as a function of progenitor radius $R$, ejecta velocity $\rm v_{s*}$, and total ejecta mass $M$, by (eqs. \ref{eq:t_br_of_vs}-\ref{eq:T_br_of_vs})
\begin{equation}
\label{eq:t_br_of_vs_Appendix}
    t_{\rm br}= 0.86 \, R_{13}^{1.26} {\rm v_{\rm s*,8.5}^{-1.13}}
(f_{\rho}M_0\kappa_{0.34})^{-0.13}\,\text{hrs},
\end{equation}
\begin{equation}
\label{eq:L_br_of_vs_Appendix}
L_{\rm br}=3.69\times10^{42} \, R_{13}^{0.78} {\rm v_{\rm s*,8.5}^{2.11}}
(f_{\rho}M_0)^{0.11} \kappa_{0.34}^{-0.89}\,{\rm erg \, s^{-1}},
\end{equation}
\begin{equation}
\label{eq:T_br_of_vs_Appendix}
T_{\rm col,br}= 8.19 \, R_{13}^{-0.32} {\rm v_{\rm s*,8.5}^{0.58}}
(f_{\rho}M_0)^{0.03} \kappa_{0.34}^{-0.22}\,{\rm eV}.
\end{equation}
Here, $R= 10^{13}\,R_{13}$~cm, $\kappa=0.34 \, \kappa_{0.34} \, \rm \, cm^2 \, g^{-1}$, ${\rm v_{\rm s*}=v_{\rm s*,8.5}} \, 10^{8.5} \, \rm cm \,  s^{-1}$, $M_0$ denotes mass in units of solar mass, and $f_\rho\simeq1$ depends on the inner structure of the envelope (see eq.~\ref{eq:rho_in}). $\rm v_{\rm s\ast}$ is related to the characteristic ejecta velocity $\rm v_\ast$ by (eq. \ref{eq:vstar})
\begin{equation}
\label{eq:Avstar_Appendix}
    {\rm v_{\rm s\ast}}\approx 1.05 f_\rho^{-0.19}v_\ast,\quad \rm v_\ast\equiv\sqrt{E/M},
\end{equation}
where $E$ is the energy deposited in the ejecta.

Our analytic approximation for the luminosity and spectra of the shock cooling emission, taking into account deviations from a blackbody spectrum, is (eq. \ref{eq:Lnu epsilon final})
\begin{equation}
\label{eq:Lnu epsilon final_Appendix}
L_{\nu} = \begin{cases}
\left[L_{\rm BB} (0.85 \, T_{\rm col})^{-m} + L_{\nu,\epsilon}^{-m}\right]^{-1/m} & h\nu<3.5 T_{\rm col} \\
1.2 \times L_{\rm BB}(0.85 R_{13}^{0.13} t_d^{-0.13} \times T_{\rm col}) & h\nu>3.5 T_{\rm col},
\end{cases}
\end{equation}
with $m=5$ and (eqs.\ref{eq:L_nu_BB_formula}, \ref{eq: epsilon prescription})
\begin{equation}
    L_{\rm BB}=L\times\pi B_{\nu}(T_{\rm col})/\sigma T_{\rm col}^{4}
    \label{eq:L_nu_BB_formula_Appendix},
\end{equation}
\begin{equation}
 L_{\nu,\epsilon}=\frac{\left(4\pi\right)^{2}}{\sqrt{3}}r_{col,\nu}^{2}\frac{\sqrt{\epsilon_{\nu}}}{1+\sqrt{\epsilon_{\nu}}}B_{\nu}(T_{col,\nu}),\quad 
  \epsilon_\nu=\frac{\kappa_{\rm ff,\nu}}{\kappa_{\rm ff,\nu}+\kappa_{\rm es}}  \label{eq: epsilon prescription _Appendix}
\end{equation}
and (eqs. \ref{eq: r_col_nu_br_notation}-\ref{eq: T_col_nu_br_notation})
\begin{equation}
    r_{\rm col,\nu} = R + 2.18 \times 10^{13} L_{\rm br,42.5}^{0.48} T_{\rm br,5}^{-1.97}
    \kappa_{0.34}^{-0.07} \tilde{t}^{0.80} \nu_{\rm eV}^{-0.08} \, \rm cm,
\end{equation}
\begin{equation}
    T_{\rm col,\nu} = 5.47 \, L_{\rm br,42.5}^{0.05} T_{\rm br,5}^{0.92}
    \kappa_{0.34}^{0.22}
    \tilde{t}^{-0.42} \nu_{\rm eV}^{0.25} \, \rm eV,
\end{equation}
\begin{equation}
    \kappa_{\rm ff} = 0.03 \, L_{\rm br,42.5}^{-0.37} T_{\rm br,5}^{0.56} 
    \kappa_{0.34}^{-0.47}
    \tilde{t}^{-0.19}
    \nu_{\rm eV}^{-1.66} \, \rm cm^2 \, g^{-1}.
\end{equation}
Here $L_{\rm br}=L_{\rm br,42.5} 10^{42.5} \rm \, erg \, s^{-1}$, $T_{col}=5 T_{\rm col,5}$ eV, and $\nu=\nu_{\rm eV}$ eV, and $R$ in terms of the break parameters is (eq. \ref{eq: R_of_br_params})
\begin{equation}
    R = 2.41\times10^{13} \, t_{\rm br,3}^{-0.1} \,L_{br,42.5}^{0.55} \,T_{br,5}^{-2.21} \,\text{cm}.
\end{equation}

A simpler approximation, that depends only on the $L$ and $T_{\rm col}$ and is slightly less accurate, is (eq. \ref{eq:Lnu epsilon final})
\begin{multline}
   L_{\nu} = \\
   \begin{cases} \frac{\pi}{\sigma}\frac{L}{T_{\rm col}^4} \left[\, \left(\frac{B_{\nu}(0.85 T_{\rm col})}{(0.85)^4} \right)^{-m} + \right.\\
        \left. \left( \frac{8}{\sqrt{3}} x^{-0.155} T_{5}^{-0.1} \frac{\sqrt{\epsilon_{\rm a}}}{1+\sqrt{\epsilon_{\rm a}}} B_\nu(1.63 \, x^{0.247} T_{\rm col}) \right)^{-m} \right]^{-1/m} & h\nu<3.5 T_{\rm col} \\
        \quad & \quad \\
        1.2\times L_{\nu,\rm BB}(1.11 L_{42.5}^{0.03} T_{5}^{0.18} \times T_{\rm col}) & h\nu>3.5T_{\rm col}
   \end{cases}
\end{multline}
where $x=h\nu/T_{\rm col}$, $T_{\rm col} = 5 \, T_{5} \, \rm eV$, and $\epsilon_{\rm a} = 0.0055 \, x^{-1.664} T_{\rm col,5}^{-1.0996}$.

Our analytic approximations are valid at (eq. \ref{eq:3Rc 1st time}-\ref{eq:t_transp}) 
\begin{equation}
\label{eq:Ats_Appendix}
   \max \left [3 R / c , t_{\rm bo} \right] < t < \min[t_{\rm 0.7 eV}, t_{\rm tr}/a],
\end{equation}
where
\begin{equation} \label{eq:tbo_Appendix}
    t_{\rm bo} = 30 \, R_{13}^{2.16} v_{\rm s*,8.5}^{-1.58} (f_\rho M_0 \kappa_{0.34})^{-0.58} \, \rm sec, 
\end{equation}
\begin{equation} \label{eq:At_0_7eV_Appendix}
    t_{0.7 \, \rm eV} = 6.86 \, R_{13}^{0.56} {\rm v_{\rm s*,8.5}^{0.16}} \kappa_{0.34}^{-0.61} (f_{\rho}M_0)^{-0.06} \rm  \, days,
\end{equation}
\begin{equation}
\label{eq:At_transp}
    \begin{split}
        \begin{aligned}
            t_{\rm tr} &= 19.5 \, \sqrt{\frac{\kappa_{0.34}M_{\rm env,0}} {\rm {v_{\rm s*,8.5}}}}\, \text{days}.
        \end{aligned}
    \end{split}
\end{equation}

\bsp	
\label{lastpage}
\end{document}